\documentstyle{article}
\catcode`\@=11
\newif\if@whiledo     \newif\if@isdrafts
\newif\if@landscape   \newif\if@titlemanual
\newif\if@smallfonts  \newif\if@ispreprint

\newcommand{\@preprintnum}{}  \newcommand{\@headtitle}{}
\newcommand{\@submited}{}
\newcommand{\MANUALTITLE}{\@titlemanualtrue}
\newcommand{\DRAFT}{\@isdraftstrue}
\newcommand{\PREPRINT}{\@ispreprinttrue}
\newcommand{\submit}[1]{\renewcommand{\@submited}{\bf{#1}}}
\newcommand{\preprint}[1]{\renewcommand{\@preprintnum}{#1}}
\newcommand{\HEADTITLE}[1]{\renewcommand{\@headtitle}{#1}}
\newcommand{\@authoraddress}{}
\newcommand{\@abstract}{}
\renewcommand{\abstract}[1]{\renewcommand{\@abstract}{#1}}

\newlength{\prepX}  \newlength{\prepY}
\newlength{\titleV} \newlength{\titlebase}
\newlength{\authorV} \newlength{\authorbase}
\newlength{\abstV}  \newlength{\abstbase}
\newlength{\labst}        \newlength{\rabst}
\newlength{\submitX} \newlength{\submitY}
\newlength{\addvspAU} \newlength{\addvspAD}

\@titlemanualfalse \@isdraftsfalse  \@ispreprintfalse

\newcommand{\todayh}{\today}
\newcommand{\@landscapemode} 
{
\input{art10.sty}
\if@smallfonts
 \def\@normalsize{\@setsize\normalsize{8.7pt}\ixpt\@ixpt
 \abovedisplayskip 8.7pt plus2pt minus4pt\belowdisplayskip
\abovedisplayskip
 \abovedisplayshortskip \z@ plus3pt\belowdisplayshortskip 4pt plus4pt
 minus3pt\let\@listi\@listI}
 \def\small{\@setsize\small{7.9pt}\viiipt\@viiipt
 \abovedisplayskip 7.9pt plus 3pt minus 4pt\belowdisplayskip
\abovedisplayskip
 \abovedisplayshortskip \z@ plus2pt
\belowdisplayshortskip 3pt plus1pt minus 1pt
 \def\@listi{\leftmargin\leftmargini
\topsep 4pt plus 2pt minus 2pt\parsep 2pt
 plus 1pt minus 1pt
 \itemsep \parsep}}
 \def\footnotesize{\@setsize\footnotesize{6.71pt}\viipt\@viipt
 \abovedisplayskip 4.24pt plus 2pt minus 3pt\belowdisplayskip
\abovedisplayskip
 \abovedisplayshortskip \z@ plus 1pt
\belowdisplayshortskip 3pt plus 1pt minus
 2pt
 \def\@listi{\leftmargin\leftmargini
\topsep 3pt plus 1pt minus 1pt\parsep 2pt
 plus 1pt minus 1pt
 \itemsep \parsep}}
 \def\scriptsize{\@setsize\scriptsize{5.65pt}\vipt\@vipt}
 \def\tiny{\@setsize\tiny{4.24pt}\vpt\@vpt}
 \def\large{\@setsize\large{9.9pt}\xiipt\@xipt}
 \def\Large{\@setsize\Large{12.7pt}\xiipt\@xiipt}
 \def\LARGE{\@setsize\LARGE{15.5pt}\xvipt\@xvipt}
 \def\huge{\@setsize\huge{17pt}\xviipt\@xviipt}
 \def\Huge{\@setsize\Huge{21.3pt}\xxpt\@xxpt}
 \normalsize
\fi 

\twocolumn \sloppy \flushbottom
\setlength{\parindent}{1em}    \setlength{\leftmargini}{2em}
\setlength{\leftmarginv}{.5em} \setlength{\leftmarginvi}{.5em}
\setlength{\evensidemargin}{-20pt} \setlength{\oddsidemargin}{-20pt}
\setlength{\marginparwidth}{48pt}
\setlength{\marginparsep}{10pt}
\setlength{\topmargin}{0em}
\setlength{\textwidth}{25.7cm}  \setlength{\textheight}{15.0cm}
\setlength{\columnsep}{1.0cm}  \setlength{\columnseprule}{0pt}
\setlength{\prepX}{1.0cm}  \setlength{\prepY}{-2.0cm}
\setlength{\titleV}{-2.6cm plus 0.5cm minus 0.5cm}
\setlength{\titlebase}{1.0em plus 0.2em minus 0.1em}
\setlength{\authorV}{2em plus 0.3em minus 0.4em}
\setlength{\authorbase}{1.3em plus 0.3em minus 0.2em}
\setlength{\abstV}{2em plus 0.5em minus 0.5em}
\setlength{\abstbase}{1.1em plus 0.3em minus 0.15em}
\setlength{\labst}{7em plus 1em minus 1em}
\setlength{\rabst}{7em plus 1em minus 1em}
\setlength{\submitX}{0cm} \setlength{\submitY}{2em}
\setlength{\addvspAU}{0em}
\setlength{\addvspAD}{0.5em plus 0.3em minus 0.3em}

} 

\newcommand{\@portratemode}
{
\input{art12.sty}
\setlength{\parindent}{1em}      \setlength{\leftmargini}{2em}
\setlength{\leftmarginv}{.5em}   \setlength{\leftmarginvi}{.5em}
\setlength{\oddsidemargin}{10pt} \setlength{\evensidemargin}{10pt}
\setlength{\marginparwidth}{48pt}\setlength{\marginparsep}{10pt}
\setlength{\topmargin}{0cm}
\setlength{\textwidth}{15.5cm} 	\setlength{\textheight}{23.0cm}

\setlength{\prepX}{0.0cm}  \setlength{\prepY}{-2.2cm}
\setlength{\titleV}{-1.0cm plus 0.5cm minus 0.5cm}
\setlength{\titlebase}{1.0em}
\setlength{\authorV}{2em plus 0.3em minus .3em}
\setlength{\authorbase}{1.3em plus 0.1em minus 0.1em}
\setlength{\abstV}{1.5em plus 0.5em minus 0.5em}
\setlength{\abstbase}{1.1em plus 0.3em minus 0.15em}
\setlength{\labst}{0em minus 0.5em}  \setlength{\rabst}{8em minus 0.5em}
\setlength{\submitX}{0cm} \setlength{\submitY}{5em plus 5em minus 5em}
\setlength{\addvspAU}{0em} \setlength{\addvspAD}{0.0em}
} 

\newcommand{\XY}[2]{\hspace{#1}\vspace{#2}}
\newcommand{\@draftT}
{
 \begin{minipage}[t]{0.3\textwidth}
 \renewcommand{\arraystretch}{1.1}
 \if@isdrafts
 \begin{flushleft}
  \begin{tabular}{l}
   {\bf \fbox{DRAFT}} \\
   Compiled at \\
   \todayh
  \end{tabular}
 \end{flushleft}
 \else
 \vspace{3em}
 \fi
 \end{minipage}
}

\newcommand{\@preprint}%
{
 \begin{minipage}[t]{0.3\textwidth}
 \if@ispreprint
 \begin{flushright}
 \renewcommand{\arraystretch}{1.1}
 \begin{tabular}{l}
 \ifx\@empty\@preprintnum\else  \@preprintnum \\  \fi
 \ifx\@epmty\@date \else \@date \fi
 \end{tabular}
 \end{flushright}
 \fi
 \end{minipage}
}

\newcommand{\@submit}{ \ifx\@empty\@submited \else \@submited \fi}

\newcommand{\@draft}{ \if@isdrafts \todayh \fi }
\renewcommand{\author}[1]
{%
   \expandafter\def\expandafter\@authoraddress\expandafter
   {\@authoraddress{\large{#1}}\\\vspace{\addvspAU}}%
}%
\newcommand{\address}[1]%
{%
 \expandafter\def\expandafter\@authoraddress\expandafter
 {\@authoraddress{\normalsize{#1}}\\ \vspace{\addvspAD}}%
}

\newcommand{\addthanks}[1]
{
  \begingroup
  \def\protect{\noexpand\protect\noexpand}\xdef\@thanks{%
  \@thanks\protect\footnotetext[\the\c@footnote]{#1}}
  \endgroup
}

\def\titlepage
{%
 \@restonecolfalse
 \if@twocolumn
   \@restonecoltrue\onecolumn
 \else
   \newpage
 \fi
 \thispagestyle{empty}\c@page\z@
}

\def\endtitlepage%
{%
 \if@restonecol\vfil\twocolumn \else \vfil\newpage\fi
}

\def\maketitle
{%
 \flushbottom
 \begin{titlepage}
     \@maketitle%
  \endgroup
  \setcounter{footnote}{0}
  \end{titlepage}
  \normalsize
  \gdef\@author{}\gdef\@title{}
  \let\@authoraddress\relax \let\address\relax \let\author\relax
  \let\@maketitle\relax      \let\maketitle\relax
  \newpage
}

\def\@maketitle%
{%
\begingroup
  \renewcommand{\baselinestretch}{1}
  \skip\footins 3pt plus 4pt minus 3pt
  \footskip 0pt
  \def\thefootnote{\fnsymbol{footnote}}
  \def\@makefnmark{\hbox to 0pt{$^{\@thefnmark}$\hss}}
  \def\thefootnote{\fnsymbol{footnote}}
  \if@landscape\else \global\@topnum\z@ \fi
  \kern\prepX\raisebox{-\prepY}{
  \begin{minipage}[t]{\textwidth}
    {\@draftT \hfill \@preprint}
  \end{minipage}
  }
  \begin{center}
  {
    \vspace{\titleV} \setlength{\baselineskip}{\titlebase}
    \huge\bf \@title\par
  }\par
  {
    \vspace{\authorV}\setlength{\baselineskip}{\authorbase}
    \par \@authoraddress
  }
  \@thanks
  \vspace{\abstV}
  \end{center}
  \centering{\large \bf Abstract}
  {\nopagebreak
  \begin{list}{}
    {\setlength{\rightmargin}{\rabst} \setlength{\leftmargin}{\labst}
         \small \setlength{\baselineskip}{\abstbase}}
    \item[] \par \protect\@abstract
  \end{list}
\ifx\@empty\@submit\else\nopagebreak[4]\XY{\submitX}{\submitY}%
\nopagebreak[4]\@submit\fi
  }
}

\long\def\ifthenelse#1#2#3{\setbox\@tempboxa
  \vbox{\let\protect\noexpand
        \let\or\@or \let\and\@and \let\not\@neg \let\(\@lparen \let\)\@rparen
        {\let\if\relax\let\ifx\relax\let\ifnum\relax\let\fi\relax
        \let\else\relax \xdef\@gtempa{#1}}\expandafter
        \@eval \ifnum \@gtempa\relax \def\@term{T}\else \def\@term{F}\fi
            \@endeval\global\let\@gtempa\@val}\if\@gtempa T#2\else#3\fi}

\def\@eval{\def\@val{F}\def\@orop{T}\def\@negate{F}}

\def\@endeval{\if\@negate T\def\@negate{F}\if\@term T\def\@term{F}\else
               \def\@term{T}\fi\fi
       \if\@orop T\if\@val F\let\@val\@term\fi
             \else \if\@val T\let\@val\@term\fi\fi}

\def\@or{\relax\def\noexpand\@term{T}\else \def\noexpand\@term{F}\fi
         \noexpand\@endeval \def\noexpand\@orop{T}\ifnum}

\def\@and{\relax\def\noexpand\@term{T}\else \def\noexpand\@term{F}\fi
          \noexpand\@endeval\def\noexpand\@orop{F}\ifnum}

\def\@neg{1=1 \fi \def\noexpand\@negate{T}\ifnum}

\def\@lparen{1=1 \fi \begingroup \noexpand\@eval \ifnum}

\def\@rparen{\relax\def\noexpand\@term{T}\else
\def\noexpand\@term{F}\fi
\noexpand\@endeval \if\noexpand\@val T\gdef\noexpand\@gtempb{1=1}\else
            \gdef\noexpand\@gtempb{1=0}\fi
            \endgroup \ifnum\noexpand\@gtempb}

\def\equal#1#2{1=1 \fi \edef\noexpand\@tempa{#1}\edef\noexpand%
\@tempe{#2}\ifx
                \noexpand\@tempa\noexpand\@tempe
                \def\noexpand\@tempa{1=1}\else
                \def\noexpand\@tempa{1=0}\fi \ifnum \noexpand\@tempa}

\long\def\whiledo#1#2{\ifthenelse{#1}{\@whiledotrue
\@whilesw\if@whiledo
\fi{#2\ifthenelse{#1}{\@whiledotrue}{\@whiledofalse}}}{}}

\typein[\MODE]{Landscap (l)  or Portlate(p)}

\ifthenelse{\equal{\MODE}{l}}%
         {  \@landscapetrue \@smallfontstrue}
         {  \@landscapefalse \@smallfontsfalse}%

\if@landscape
  \@landscapemode
\else
  \@portratemode
\fi
\@afterindenttrue
\newcommand{\Gp}{\gamma^5}
\newcommand{\VEV}[1]{\langle{#1}\rangle}
\newcommand{\xbar}[1]{{{#1}\kern-0.5em\hbox{$/$}}}

\newcommand{\del}[1]{\partial_{#1}}
\newcommand{\delu}[1]{\partial^{#1}}

\newcommand{\DISP}{\displaystyle}
\newcounter{subequation}

\newcommand{\psib}{\bar{\psi}}

\newcommand{\tr}{\mbox{tr}}
\newcommand{\Ref}[1]{(\ref{#1})}

\renewcommand{\theequation}{{\arabic{section}.\arabic{equation}}}%

\newenvironment{Eqnarray}%
{%
 \setcounter{subequation}{\value{equation}}%
 \setcounter{equation}{0}%
 \addtocounter{subequation}{1}
 \renewcommand{\theequation}%
 {{\arabic{section}.\arabic{subequation}\alph{equation}}}%
 \begin{eqnarray}
}{%
 \end{eqnarray}
 \setcounter{equation}{\value{subequation}}%
 \renewcommand{\theequation}{{\arabic{section}.\arabic{equation}}}%
 \hspace{-\parindent}
}

\newcommand{\gtrsim}{\mathop{>}\limits_{\displaystyle{\sim}}}
\newcommand{\lessim}{\mathop{<}\limits_{\displaystyle{\sim}}}

\def\section{%
 \setcounter{equation}{0}
 \@startsection {section}{1}{\z@}{3.5ex plus 1ex minus
 .2ex}{2.3ex plus .2ex}{\Large\bf}
 }
\def\subsection{%
\@startsection{subsection}{2}{\z@}{3.25ex plus 1ex minus
 .2ex}{1.5ex plus .2ex}{\large\bf}}
\def\subsubsection{\@startsection{subsubsection}{3}{\z@}{3.25ex plus
1ex minus .2ex}{1.5ex plus .2ex}{\normalsize\bf}}
\def\paragraph{\@startsection
 {paragraph}{4}{\z@}{3.25ex plus 1ex minus .2ex}{1em}{\normalsize\bf}}
\def\subparagraph{\@startsection
 {subparagraph}{4}{\parindent}{3.25ex plus 1ex minus
 .2ex}{1em}{\normalsize\bf}}

\def\baselinestretch{2}

\def\@setsize#1#2#3#4{\@nomath#1%
   \let\@currsize#1\baselineskip
   #2\baselineskip\baselinestretch\baselineskip
   \parskip\baselinestretch\parskip
   \setbox\strutbox\hbox{\vrule height.7\baselineskip
      depth.3\baselineskip width\z@}%
   \normalbaselineskip\baselineskip#3#4}

\skip\footins 20pt plus4pt minus4pt

\def\@xfloat#1[#2]{\ifhmode \@bsphack\@floatpenalty -\@Mii\else
   \@floatpenalty-\@Miii\fi\def\@captype{#1}\ifinner
      \@parmoderr\@floatpenalty\z@
    \else\@next\@currbox\@freelist{\@tempcnta\csname ftype@#1\endcsname
       \multiply\@tempcnta\@xxxii\advance\@tempcnta\sixt@@n
       \@tfor \@tempa :=#2\do
                        {\if\@tempa h\advance\@tempcnta \@ne\fi
                         \if\@tempa t\advance\@tempcnta \tw@\fi
                         \if\@tempa b\advance\@tempcnta 4\relax\fi
                         \if\@tempa p\advance\@tempcnta 8\relax\fi
         }\global\count\@currbox\@tempcnta}\@fltovf\fi
    \global\setbox\@currbox\vbox\bgroup
    \def\baselinestretch{1}\small\normalsize
    \boxmaxdepth\z@
    \hsize\columnwidth \@parboxrestore}
\long\def\@footnotetext#1{%
\insert\footins{\def\baselinestretch{1}\footnotesize
    \interlinepenalty\interfootnotelinepenalty
    \splittopskip\footnotesep
    \splitmaxdepth \dp\strutbox \floatingpenalty \@MM
    \hsize\columnwidth \@parboxrestore
   \edef\@currentlabel{%
\csname p@footnote\endcsname\@thefnmark}\@makefntext
    {\rule{\z@}{\footnotesep}\ignorespaces
      #1\strut}}}

\def\singlespace{%
\vskip\parskip%
\vskip\baselineskip%
\def\baselinestretch{1}%
\ifx\@currsize\normalsize\@normalsize\else\@currsize\fi%
\vskip-\parskip%
\vskip-\baselineskip%
}

\def\spacing#1{\par%
 \def\baselinestretch{#1}%
 \ifx\@currsize\normalsize\@normalsize\else\@currsize\fi}

\everydisplay{
   \abovedisplayskip \baselinestretch\abovedisplayskip%
   \belowdisplayskip \abovedisplayskip%
   \abovedisplayshortskip \baselinestretch\abovedisplayshortskip%
   \belowdisplayshortskip  \baselinestretch\belowdisplayshortskip}

\PREPRINT

\date{November 1993}
\preprint{DPNU-93-45\\ CHIBA-EP-73\\KEK-TH-380\\KEK Preprint 93-171}

\title{
 Phase Structure \\of\\ the Gauged Yukawa Model
 \thanks{
    \protect\parbox[t]{\textwidth}{
    Work supported in part by the Grant-in-Aid
    for Scientific Research from the Ministry of Education,
    Science and Culture (No. 05640339),\protect \\
    \protect\hspace{5em} the Ishida Foundation and the International
    Collaboration Program of the
    Japan Society for the Promotion of Science.
    }}
 }
\author{ {\sc Kei-ichi Kondo}%
\thanks{e-mail: kondo@tansei.cc.u-tokyo.ac.jp;
        kondo@cuphd.nd.chiba-u.ac.jp}}
\address{Department of Physics, Faculty of Science \\
        $\&$ Graduate School of Science and Technology,%
         \\
         Chiba University, Chiba 263, Japan}
\author{ {\sc Akihiro Shibata}%
     \thanks{e-mail: %
          shibata@eken.phys.nagoya-u.ac.jp;
          d42837a@nucc.cc.nagoya-u.ac.jp }
       }
\address{Department of Physics, Nagoya University\\
         Nagoya 464-01, Japan }
\author{ {\sc Masaharu Tanabashi}
  \thanks{
    e-mail: tanabash@theory.kek.jp
  }
}
\address{
 National Laboratory for High Energy Physics (KEK) \\
  Tsukuba, Ibaraki 305, Japan }
\author{ \large and }
\author{{\sc Koichi Yamawaki}%
      \thanks{e-mail: yamawaki@eken.phys.nagoya-u.ac.jp;
                       b42060a@nucc.cc.nagoya-u.ac.jp}
}

\address{Department of Physics, Nagoya University\\
         Nagoya 464-01, Japan }

\abstract
{
    Based on the ladder Schwinger-Dyson equation,
    we investigate phase structure of the gauged Yukawa model
    possessing a global $SU(2)_L \times SU(2)_R$
    symmetry and an unbroken (vector-like) gauge symmetry.
    We show that even when we tune the squared mass of
    the scalar boson
    in the Lagrangian to be positive, there still exists
    the dynamical chiral symmetry breaking
    due to fermion pair condensate (VEV of the composite scalar)
    triggered
    by the strong Yukawa coupling larger than a certain critical value.
    We find a ``nontrivial ultraviolet fixed line'' and ``renormalized
    trajectories'' in the three-dimensional coupling space of
    the Yukawa coupling, the gauge coupling and the
    ``hopping parameter'' of
    the elementary scalar field.
    Presence of the gauge coupling is crucial to existence
    of the fixed line.
    Implications of the result for the lattice calculation and
    the top quark condensate model are  also discussed.
    \nopagebreak[4]
    \par\nopagebreak[4]
    \vspace{3em plus 1em minus 2em} \nopagebreak[4]
    {\bf Submitted to Progress of Theoretical Physics.}
    \vspace{-3.0em}
}
\begin{document}
\spacing{1.0}
\maketitle
\spacing{1.6}

\section{Introduction}
    As it turned out, the standard model of modern particle physics is
extremely successful. However, the central mystery of this model is
the two missing ingredients, the Higgs boson and the top quark: Why
are they so heavy?  The ever increasing experimental bound of the top
quark mass is now getting closer to the weak scale 246 GeV (the present
LEP constraint is $164 {\rm GeV} \pm 27 {\rm GeV}$, while the CDF
bound is $> 113 {\rm GeV}$ \cite{kn:Holl93}).
 This seems to suggest a special role of the top
quark in the electroweak symmetry breaking and hence a strong
connection with yet another missing ingredient, the Higgs boson.  In
contrast to the passive role of Yukawa couplings simply picking up the
already tuned Higgs vacuum expectation value (VEV) to give mass to the
known fermions, such a strong Yukawa coupling of the top quark may
affect the entire dynamical picture (phase structure) of the standard
Higgs sector.
\par

   This situation can be most naturally understood by the Top Quark
Condensate Model (Top Mode Standard Model)\cite{kn:Yama92} which was
proposed by Miransky, Tanabashi and Yamawaki (MTY)\cite{kn:MTY89} and
by Nambu\cite{kn:Namb89} independently.  This model entirely replaces
the standard Higgs doublet by the composite one formed by the strongly
coupled short range dynamics (four-fermion interaction) which is
responsible for the top quark condensate.  The Higgs boson emerges as
a $\bar t t$ bound state and hence is deeply connected with the top
quark itself. Actually, based on the explicit solution of the ladder
Schwinger-Dyson (SD) equation of the gauged Nambu-Jona-Lasinio (NJL)
model (QCD plus four-fermion interaction), MTY\cite{kn:MTY89}
predicted the top quark mass to be about $250 {\rm GeV}$
(for the Planck
scale cutoff), which is just the order of the electroweak symmetry
breaking scale.  This model was further formulated in an elegant
fashion by Bardeen, Hill and Lindner\cite{kn:BHL90} in the
standard model language: They incorporated the composite
Higgs loop effects, which turned out to reduce the above MTY value
down to $220 {\rm GeV}$, a somewhat smaller value but still on the
order of the weak scale.  Although the prediction appears to be
substantially higher than the present experimental bound mentioned
above, there still remains a
possibility that (at least) the essential feature of the top quark
condensate idea may eventually survive.
\par

  What is the origin of the top mode four-fermion interaction, then?
This question was first addressed in a concrete manner by Kondo,
Tanabashi and Yamawaki (KTY)\cite{kn:KTY90}
(See also Ref.\cite{kn:KTY93}), who
  suggested a heavy
spinless boson exchange model with the strong Yukawa coupling
(``Yukawa-driven Top Mode Model (YTMM)'') in the framework of the SD
equation.  The idea was further discussed by Clague and
Ross\cite{kn:CR91} in a slightly different framework.  One might be
tempted to consider an alternative, the heavy spin 1 boson exchange
model.  However, as was already pointed out\cite{kn:KTY90},
 this kind of model does
not give rise to the desired four-fermion interaction, $g^{(2)}$ term
in Ref.\cite{kn:MTY89}, which communicates the top quark condensate
to the bottom quark mass. Hence it has no chance to give mass to the
bottom quark without suffering from the axion problem (See the second
paper of Ref.\cite{kn:MTY89}, and also
Refs.\cite{kn:Yama92,kn:Tana92,kn:EK93}).
This difficulty also applies to the more recently proposed
models\cite{kn:NNT91,kn:Hill91} of heavy spin 1 boson exchange.
\par

   In YTMM\cite{kn:KTY90,kn:KTY93,kn:CR91} the top quark condensate
is triggered by the strong Yukawa interaction.  Even when we start
with the same Lagrangian as the standard model (with opposite sign of
the squared scalar mass and hence without Higgs VEV), we have a quite
different dynamical picture: The electroweak symmetry breaking takes
place mainly due to the top quark condensate instead of the ad hoc
tuned Higgs VEV.  Thus the principal role of the ``elementary'' Higgs
is not to break the electroweak symmetry through its VEV but to supply
a strong attractive force between the top and anti-top through the
Yukawa coupling so as to trigger the top quark condensate or the
composite Higgs VEV.\footnote{
This picture is contrasted with the Nambu's
bootstrap\cite{kn:Namb88}
which implies identification of
the ``elementary'' Higgs with the composite one, while we here
distinguish between the two Higgses, i.e., one (elementary) with the
GUT or Planck scale mass and the other (composite) with the weak scale
mass whose VEV makes a dominant contribution to the $W/Z$ masses.
}

   It should be noted that YTMM\cite{kn:KTY90,kn:KTY93,kn:CR91} yields
potentially strong higher dimensional operators in addition to the
original four-fermion interaction\cite{kn:MTY89,kn:BHL90} relevant to
the top quark condensate.  Actually, it was first pointed out by
Suzuki\cite{kn:Suzu90} that inclusion of such higher dimensional
operators may reduce the top mass prediction of the top mode standard
model.  The role of higher dimensional operators was further
clarified by Hasenfratz et al.\cite{kn:HHJKS91} (see also
Ref.\cite{kn:Zinn91}). Once we specify a possible underlying
theory, we can estimate the effects of these higher dimensional
operators on the top mass prediction.  For example, the spin 1 boson
exchange model does not seem to yield large coefficients for the
higher dimensional operators \cite{kn:Hill91}.  If, on the other hand,
there exist large effects of the higher dimensional operators in YTMM,
then this model may have a chance to predict a smaller top mass to be
consistent with even the present LEP experiments mentioned above.
\footnote{
Although the YTMM might then loose predictive power for the top
mass itself, it might maintain predictability of the top to Higgs mass
ratio which would be testable in the future experiments.}
\par

Apart from the top quark condensate, the strongly coupled Yukawa model
may be applied to other models beyond the standard model.
An interesting
example is a ``Heavy Scalar Technicolor Model''\cite{kn:Simm89}
in which the role of the ETC gauge
bosons in the ordinary technicolor scenario is replaced by
 the heavy weak doublet scalar boson exchange which communicates
the technifermion condensate to the ordinary fermion mass.
This is actually described by the gauged Yukawa model, with the gauge
interaction being the technicolor instead of the QCD.
Such a model may have
a ``walking/standing technicolor''\cite{kn:Hold85}
gauge coupling.
 When the Yukawa coupling becomes strong, the dynamical feature
of this model is expected to be somewhat similar to the
 ``strong ETC model''\cite{kn:MY89}
 based on the gauged NJL model, but may provide a rather different
 picture to be testable in the future experiments.
 \par

   In order to draw a definite conclusion on the above problems,
   however, we need to solve nonperturbative dynamics of
   the strong Yukawa coupling. This is a very difficult task,
   however, and
   cannot be done at once.
   Actually, as the first step KTY\cite{kn:KTY90}
   studied the phase structure of the pure Yukawa theory (without
   gauge coupling) with a global $SU(2)_L \times SU(2)_R$ symmetry
   within
   the framework of the ladder SD equation.
    (The extension has also been made\cite{kn:KTY93}
     to include the $SU(2)_L \times U(1)_Y$-invariant
    Yukawa model, which corresponds to
     the realistic cases of the
    top quark condensate and the heavy scalar technicolor models.)
    It was shown that without tadpole,
    we have a clear signal of  the dynamical
    symmetry breaking;
    a vanishing scalar VEV and nonzero fermionic condensate
    (fermion dynamical mass $M \not= 0$) for the strong
    Yukawa coupling larger than the critical coupling.
    However, inclusion
    of the tadpole correlates the both order parameters and makes the
    concept of ``dynamical symmetry breaking'' somewhat ambiguous,
    even if we have a nonzero fermionic condensate
    for the strong coupling region
    (Actually, there exists a critical coupling.). KTY
         proposed a possible criterion for the
     ``dynamical symmetry breaking''
    that the fermionic current contribution dominates the scalar one
    to the decay constant $F_\pi$ of the Nambu-Goldstone (NG) bosons
    in such a way that they are mostly the composite Higgs with a
    small admixture of the elementary one. Such a situation in fact
    turned out to be the case in a wide
      range of the parameter space in the $SU(2)_L \times SU(2)_R$
      Yukawa model.
\par
   In this paper we shall extend the previous analysis\cite{kn:KTY90}
   so as to include a vector-like gauge coupling
   (``gauged Yukawa model''
   in the same sense as the gauged NJL model) and an analog of
   the hopping parameter
   $Z_\phi$ of the elementary scalar field $\phi$
   whose kinetic term is parameterized as
   $\frac{Z_\phi}{2} (\partial_\mu\phi)^2$.
   Based on the SD equation with the standing gauge coupling (running
   effect ignored), we study the phase structure of the model
    in the three-dimensional parameter space
   ( Yukawa coupling, gauge coupling, $Z_\phi$ )
   compared with the previous analysis in
   one-dimensional space of the Yukawa coupling\cite{kn:KTY90}.
   \par
  Inclusion of the ``hopping parameter'' $Z_\phi$
  is of course vital to our
  analysis, since this is the very parameter
  that characterizes the deviation,
  though not all,
  from the gauged NJL model\footnote{
   For extensive study of the gauged NJL model
   with standing gauge coupling, see Ref.\cite{kn:KTY93a}
   and references
   cited therein. As to the model with running coupling,
    see Ref.\cite{kn:KSY91}.
   }
   ($Z_\phi = 0$) and hence from
   the original top quark condensate model.
 Actually, we find that the fermion mass is lowered due to $Z_\phi >0$
   when compared with the
   gauged NJL model.
\par
  Inclusion of the gauge coupling is motivated by the
  realistic situation of the top quark condensate model where the QCD
   coupling makes a significant
  contribution to the top mass prediction \cite{kn:Yama92}. Actually,
  the gauge coupling drastically changes the phase structure:
  We discover the ``fixed line'' and the ``renormalized trajectories''
   which can only be revealed by the presence of the (standing)
   gauge coupling.
   This we believe is a novel feature of the present model
   and is one of the major achievements of this paper.
    Roughly speaking, the renormalized trajectories
   correspond to the boundary which separates the region of
   the ``dynamical symmetry breaking'' (fermionic dominance)
   from that of the
   ``non-dynamical symmetry breaking'' (fermionic non-dominance)
   in the sense of
   Ref.\cite{kn:KTY90} mentioned above.

The paper is organized as follows.
In Section 2 we set up our machinery,
the ladder SD equation for the gauged Yukawa model with a
standing gauge coupling and a global $SU(2)_L \times SU(2)_R$
symmetry.
We include a tadpole contribution.
In Section 3 we present a new formula
(``generalized Pagels-Stokar (PS) formula'') for
the decay constant $F_\pi$ of the NG boson in this model.
$F_\pi^2$ consists of two parts
one from the scalar current, and the other from the fermionic one,
the latter being given by the PS formula\cite{kn:PS79}
originally developed
in the theory without scalar field.
The very structure of our formula dictates that the fermion mass must
be smaller in the gauged Yukawa model than in the gauged NJL model.
Then in Section 4 numerical analysis of the phase structure of this
model is given.
We identify
the line of constant $M$ (fermion dynamical mass)
and constant $F_\pi$ with
the ``renormalization-group (RG) flow''.
We discover a ``fixed line'' and ``renormalized trajectories''
in the three-dimensional parameter space
of (Yukawa coupling, gauge coupling,
$Z_\phi$). Section 5 is the analytical study of the model
which reproduces the numerical one in Section 4 and further provides
the scaling relation near the phase transition points
(critical surface).
We find an analytical expression for the whole fixed line.
Section 6 is devoted to the conclusion
and discussions. Possible implications
for the lattice studies and the top quark condensate model
are discussed.
It is suggested that the top quark mass prediction of YTMM
may be lower than
the original top mode prediction\cite{kn:MTY89,kn:BHL90}
 based on the gauged NJL model.


\section{Schwinger-Dyson equation}
We start with the Lagrangian of the gauged Yukawa model (Yukawa model
coupled to a vector-like $SU(N_c)$ gauge interaction) with a global
$SU(2)_L\times SU(2)_R$ symmetry:
\begin{eqnarray}
{\cal L}
  &=& \psib \left(
         i\xbar{\partial} + \frac{g}{\sqrt{N_c}} \xbar{G}
      \right)\psi
     -\tilde g_y \left[
        \psib_{L} \Phi^{\dagger} \psi_{R} + \psib_{R} \Phi \psi_{L}
      \right]
     -\frac{1}{4}  G^{\alpha}_{\mu\nu}G^{\alpha \mu\nu}
  \nonumber\\
  & &+\frac{1}{4}\tr (\del{\mu} \Phi^{\dagger} \delu{\mu}\Phi )
     -\frac{m_\phi^2 }{4} \tr(\Phi^{\dagger} \Phi )
     - \frac{\tilde \lambda_{\phi}}{16}
         \left(\tr( \Phi^{\dagger} \Phi ) \right)^2,
\label{eq:Lagrangian0}
\end{eqnarray}
where $\Phi$ is a $2\times2$ matrix, $\Phi=\sigma +i\tau^a\pi^a$
with Pauli matrices $\tau^a (a=1,2,3)$,
$\psi$ is a fermion doublet field, $G_\mu:=G_\mu^{\alpha}T^{\alpha}
(\alpha=1,\dots,N_c^2-1)$ are $SU(N_c)$ gauge fields, and
$\psi_{L/R}:= P_{L/R} \psi$, with $P_{L/R} := (1 \mp \gamma^5)/2$
being chiral projection operators.
We take $m_\phi^2 >0$.
\par

It is well known that this system with negative $m_\phi^2$ causes the
spontaneous chiral symmetry breaking already
 at the tree level and the
fermion acquires its mass in a passive manner from the VEV of the
scalar field.
However, this picture is not appropriate
for the strong coupling Yukawa region
where the feedback from the fermion sector to the scalar potential is
significant as stressed before.
In this region  we expect the fermion determinant plays a more active
role in the spontaneous chiral symmetry breaking and we need to treat
the dynamics in a  nonperturbative manner.
Such a nonperturbative dynamics may  be investigated
by using a truncated
set (``ladder'')
of the Schwinger-Dyson (SD) self-consistency equations as the first
approximation.
\par

Since a  concept of  renormalized parameters becomes somewhat
obscure in this kind of nonperturbative analysis due to the ambiguity
of renormalization schemes, we study the physical quantities (e.g.,
the fermion dynamical mass $M$, the NG boson decay
constant $F_\pi$, etc.)  directly as functions of bare parameters
(e.g., $m_\phi$, $\tilde g_y$, etc.) and the ultraviolet (UV)
cutoff $\Lambda$.
Once these physical quantities are calculated,
we can determine the scaling properties
of those bare parameters (``renormalization group (RG)'' )
in such a way
as to fix the physical quantities on the variation of the cutoff
$\Lambda$.
Of course the RG functions determined in this
manner are not unique and depend on the choice of the ``physical
parameters'' utilized in this procedure.  It should be emphasized,
however,
that the qualitative feature of the phase diagram is expected not
to change depending upon the choice of the parameters,
as far as such a
choice covers (at least) minimal set of
possible relevant operators.
In this paper, we choose the dynamical mass of the fermion $M$ and the
NG boson decay constant $F_\pi$ for such physical parameters.

Now, we turn to the ladder SD equation of the gauged
Yukawa model.
It is convenient to rewrite the Lagrangian Eq.\Ref{eq:Lagrangian0}
into
\begin{eqnarray}
{\cal L}
   &=& \psib \left(
         i\xbar{\partial} + \frac{g}{\sqrt{N_c}} \xbar{G}
       \right)\psi
       - g_y \left[
           \psib_{L} \Phi^{\dagger} \psi_{R} + \psib_{R} \Phi \psi_{L}
         \right]
      - \frac{1}{4}  G^{\alpha}_{\mu\nu}G^{\alpha \mu\nu}
   \nonumber\\
   & &+ \frac{Z_{\phi} N_c}{4} \tr \left(
          \del{\mu} \Phi^{\dagger} \delu{\mu}\Phi
        \right)
      - \frac{M_\phi^2 N_c }{4} \tr \left( \Phi^{\dagger} \Phi \right)
      - \frac{\lambda_{\phi} N_c}{16}
           \left( \tr(\Phi^{\dagger} \Phi) \right)^2,
\label{eq:Lagrangian}
\end{eqnarray}
where we have rescaled the scalar field
$\Phi$: $\Phi\rightarrow \sqrt{Z_\phi N_c}\Phi$ and defined
parameters: $g_y^2 := N_c Z_\phi \tilde g_y^2$,
$M_\phi^2 := Z_\phi m_\phi^2$ and
$\lambda_\phi := N_c Z_\phi^2 \tilde \lambda_\phi$.
  Again note that we take  $M_\phi^2>0$,
since we are interested in the
symmetry breaking due to the strong Yukawa coupling.
Here we take  $Z_\phi > 0$ in order to guarantee
the stability of the vacuum.
 $Z_\phi$ plays a role of the ``hopping parameter''
in the lattice formulation (See the discussion in Section 6).
It should be noted here that our new parameterization of the gauged
Yukawa model, Eq.\Ref{eq:Lagrangian}, contains a redundant parameter
compared with the original one Eq.\Ref{eq:Lagrangian0}.
We shall return to this point later.
It should also be emphasized that the Eq.\Ref{eq:Lagrangian} is
appropriate for the comparison to the previous analysis of the gauged
NJL model ($Z_\phi=\lambda_\phi=0$).

The SD equations for the VEV of the scalar field $\VEV{\sigma}$ and
the fermion propagator $S(p)$ are given by\cite{kn:KTY90}
\begin{equation}
M_\phi^2 \VEV{\sigma} +\lambda_\phi \VEV{\sigma(\sigma^2+\vec{\pi}^2)}
    = -\frac{g_y}{N_c}\VEV{\psib\psi}
\label{eq:VEVsigma}
\end{equation}
and
\begin{eqnarray}
iS^{-1}(p)
   &=& \xbar{p}- g_y \VEV{\sigma}
      +\frac{g}{\sqrt{N_c}}\int\frac{d^4k}{(2\pi)^4}
        \gamma^{\mu} T^{\alpha} S(p-k)
                \Gamma_\alpha^{\nu}(p-k,p) D_{\mu\nu}(k)
   \nonumber \\
   & &+ g_y \int \frac{d^4k}{(2\pi)^4}
        D^\sigma(k)S(p-k)\Gamma^\sigma(p-k,p)
   \nonumber\\
   & &+ g_y \int \frac{d^4k}{(2\pi)^4}
        D^\pi_{ab}(k) i \gamma^5 \tau^b S(p-k)\Gamma^{\pi a}(p-k,p),
\label{eq:SDeqfermion}
\end{eqnarray}
respectively, with $D_{\mu\nu}(k)$ being the gauge filed propagator,
$D^\sigma(k)$ and  $D^\pi_{ab}(k) $ the scalar and the pseudoscalar
propagators, $\Gamma_\alpha^{\nu}$, $\Gamma^\sigma$
and  $\Gamma^{\pi a}$
the vertex functions.
It is hopeless to solve a  full set of the SD equations
because of the
lack of knowledge of the scalar/gauge boson propagators and the
vertex functions.
Here we take the ladder approximation in the sense that
we take bare
vertices and  bare boson propagators instead of full vertices
and full propagators:
\begin{Eqnarray}
    N_c D^{\sigma}(k) &=& \frac{i}{Z_\phi k^2-M_\phi^2}, \qquad
    \Gamma^{\sigma} = -i g_y,\\
    N_c D^{\pi}_{ab}(k)  &=&
             \frac{i}{Z_\phi k^2-M_\phi^2 }\delta_{ab},
    \qquad
    \Gamma^{\pi a} = -i g_y i\Gp \tau^a  ,\\
    N_c D_{\mu\nu}(k) &=&
      \frac{-i}{k^2} (g_{\mu\nu}-(1-\xi)\frac{k_\mu k_\nu}{k^2} ),
    \qquad
    \Gamma_\alpha^{\mu} = -i \frac{g}{\sqrt{N_c}}
                                 \gamma^{\mu} T_\alpha,
\end{Eqnarray}
with $\xi$ being the gauge fixing parameter.
Though there is no solid reason to justify this approximation,
the ladder
approximation becomes plausible, when the gauge coupling runs
slowly (``walking'')
or does not run at all (``standing'') \cite{kn:Hold85},
and the effects of the
scalar/pseudoscalar boson propagators
are small.%
\footnote{
   The effects of the scalar/pseudoscalar
   boson propagators (``rainbow'' graph )
   are actually suppressed by $1/N_c$ compared with
   those of the gauge boson propagators (gauge boson ``rainbow'')
   and the VEV of the scalar field (``tadpole'').
   (See Eqs.\Ref{eq:KernelB}--\Ref{eq:KernelA}.)
   The $N_c \to \infty$ limit yields qualitatively the same
   phase structure as $N_c =1$ case
   (see Section 5).
}
Actually, the ladder approximation
yields a successful phenomenology even for the QCD running coupling
used in the SD equation.\cite{kn:AKM91}
\par
We substitute
\begin{equation}
   i S(p)^{-1}   =  A(-p^2) \xbar{p} -  B(-p^2)
\label{eq:fermionpropagator}
\end{equation}
into Eq.\Ref{eq:SDeqfermion} and introduce the UV cutoff
$\Lambda^2$ for $p^2$ after Wick rotation.
In the ladder
approximation, we can perform the angular integration of
Eq.\Ref{eq:SDeqfermion}, which yields
(see, e.g., Ref.\cite{kn:Kond92})
\begin{Eqnarray}
  B(x)
  &=& g_y \VEV{\sigma} + \int_{0}^{\Lambda^2} dy
      {\cal K}_B(x,y) \frac{yB(y)}{A(y)^2y+B(y)^2} ,
  \label{eq:SDeq}\\
  A(x) &=& 1 + \int_{0}^{\Lambda^2} dy
      {\cal K}_A(x,y) \frac{A(y)}{A(y)^2y+B(y)^2}.
\end{Eqnarray}
The integral kernels are given by
\begin{Eqnarray}
 {\cal K}_B(x,y)
  &=& \frac{h}{N_c Z_\phi} K_B(x,y;\frac{M_\phi^2}{Z_\phi})
    + \lambda(1+\xi/3) K_B(x,y;0),
  \label{eq:KernelB} \\
 {\cal K}_A(x,y)
  &=& \frac{y}{x} \left\{
        \frac{h}{N_c Z_\phi}K_A(x,y;\frac{M_\phi^2}{Z_\phi})
      + \frac{2}{3} \xi\lambda K_A(x,y;0)
      \right\}\;,
  \label{eq:KernelA}
\end{Eqnarray}
where
\begin{Eqnarray}
       K_B(x,y;m^2) &:=& \frac{2}{\pi} \int_{0}^{\pi}d\theta
                    \frac{\sin^2\theta}{x+y-2\sqrt{xy}\cos\theta+m^2}
            \nonumber \\
              &=&  \frac{2}{ x+y+m^2 + \sqrt{ (x+y+m^2)^2 -4xy}}\;,\\
        K_A(x,y;m^2) &:=& \frac{4}{\pi}\int_{0}^{\pi}d\theta
                    \frac{\sqrt{xy}\cos\theta\sin^2\theta}
                    {x+y-2\sqrt{xy}\cos\theta +m^2} \nonumber\\
                 &=&\frac{4xy}
                    {\left[x+y+m^2+\sqrt{(x+y+m^2)^2-4xy}\right]^2}\;,
\end{Eqnarray}
and  $x$, $y$,  $h$ and $\lambda$ are defined as
$x:=-p^2$, $y:=-q^2$,
\begin{equation}
  h :=\DISP \frac{g_y^2}{8\pi^2}, \quad
  \DISP \lambda :=\frac{3g^2}{16\pi^2} \frac{C_F}{N_c},
  \label{eq:couplings}
\end{equation}
with $C_F := \sum_{\alpha} T^\alpha T^\alpha=(N_c^2-1)/2N_c$ being the
quadratic Casimir of the fermion representation.
\par

The VEV of the scalar field $\VEV{\sigma}$ is determined from
Eq.\Ref{eq:VEVsigma} under factorizability assumption,
which reads\cite{kn:KTY90}
\begin{equation}
\VEV{\sigma}
    = -\frac{g_y}{N_c}\frac{\VEV{\psib\psi}}{M_\phi^2}
      +\lambda_\phi \frac{g_y^3}{N_c^3 M_\phi^8}
      (\VEV{\bar\psi\psi})^3
      +\cdots,
\label{eq:oderpara}
\end{equation}
where the fermion condensate $\VEV{\psib\psi}$ is given by
\begin{equation}
    \VEV{\psib\psi} = -\int \frac{d^4q}{(2\pi)^4}\tr S(q)
    = -\frac{N_c}{2\pi^2}\int_0^{\Lambda^2}\!\!\! dy
                           \frac{yB(y)}{A^2(y)y+B^2(y)}.
    \label{eq:condensation}
\end{equation}
At this stage, the
VEV $\VEV{\sigma}$, the chiral condensate $\VEV{\psib\psi}$ and the
fermion mass function $B(x)$ are inter-related among each other
and are
responsible for the chiral symmetry breaking.
The ${\cal O}(\lambda_\phi)$ or higher terms in Eq.(\ref{eq:oderpara})
do not produce significant effect near the critical point
$|\VEV{\bar\psi\psi}|\ll M_\phi^3$.
We thus disregard those effects in the following calculations.
Then
$\VEV{\sigma}$ is given by the ``tadpole'' contribution alone:
\begin{equation}
   \VEV{\sigma}
    = -\frac{g_y}{N_c}\frac{\VEV{\psib\psi}}{M_\phi^2}
    = \frac{g_y}{2\pi^2 M_\phi^2}
        \int_0^{\Lambda^2}\!\!\! dy
                           \frac{yB(y)}{A^2(y)y+B^2(y)}.
\label{eq:tadpole}
\end{equation}

  Combining Eq.\Ref{eq:SDeq} and Eq.\Ref{eq:tadpole},
 we finally obtain the closed SD equation for
the fermion propagator:
\begin{Eqnarray}
        B(x) &=& \frac{4h}{M_\phi^2}
           \int_{0}^{\Lambda^2}\!\!\! dy \frac{yB(y)}{yA(y)^2+B(y)^2}
           +\int_{0}^{\Lambda^2}\!\!\!dy
                 {\cal K}_B(x,y) \frac{yB(y)}{yA(y)^2+B(y)^2} \;,
          \nonumber\\\label{eq:SDeqB}\\
        A(x) &=& 1
                + \int_0^{\Lambda^2}\!\!\!
                dy {\cal K}_A(x,y) \frac{A(y)}{yA(y)^2+B(y)^2}
               \label{eq:SDeqA} \;.
\end{Eqnarray}
\par

In this paper we take the Landau gauge ($\xi=0$).  The Landau gauge is
required to be consistent with the bare vertex function approximation
at least in the pure gauge theory due to the Ward-Takahashi identity,
since the part of the kernel ${\cal K}_{A}$ from the gauge interaction
is identically zero in this gauge and then the SD equation reduces to
$A(-p^2) \equiv 1$ (see, e.g., Ref.\cite{kn:Kond92}). For simplicity of
calculation we take a further approximation
for the integral kernel from
the Yukawa interaction such that ${\cal K}_{A} = 0$.  For this
approximation to be consistent with the bare vertex approximation, the
solution of the coupled SD equations must lead to the result: $A(-p^2)
\cong 1$ for $A(-p^2)$, namely, the deviation of $A(-p^2)$ from 1 must
be small.  Indeed this has been confirmed by the previous
work\cite{kn:KTY90}
on the Yukawa model and also in the massive vector boson
model\cite{kn:Kond89}.  Thus in what follows we put
\begin{equation}
  A(-p^2) \equiv 1\;,
\end{equation}
which implies no wave function renormalization for the fermion.  Then
we only have to solve the single integral equation for the fermion
mass function $B(x)$:
\begin{equation}
  B(x) = \frac{4h}{M_\phi^2}
             \int_{0}^{\Lambda^2}\!\!\! dy \frac{yB(y)}{y+B(y)^2}
            +\int_{0}^{\Lambda^2}\!\!\!dy
                 {\cal K}_B(x,y) \frac{yB(y)}{y+B(y)^2}.
\label{eq:SDeqBB}
\end{equation}
This is our basic equation.
\par
  Although Eq.\Ref{eq:SDeqBB} can be solved numerically,
here we solve it by converting the integral equation
into a more tractable differential equation plus boundary conditions.
This can be done by
adopting an approximation\cite{kn:LAK}
for the kernel:
\begin{eqnarray}
    K_B(x,y;m^2) \simeq \theta(x-y)\frac{1}{x+m^2}
                  + \theta(y-x)\frac{1}{y+m^2} \; ,
\end{eqnarray}
where $\theta(x)$ is the step function.
Then
the SD equation Eq.\Ref{eq:SDeqB} is reduced to
\begin{eqnarray}
        \lefteqn{
            B(x) = \frac{4h}{M_\phi^2}
             \int_{0}^{\Lambda^2}\!\!\! dy \frac{yB(y)}{y+B(y)^2}
         }
         \nonumber\\
         &+&\int_{0}^{\Lambda^2}\!\!\!dy
             \left[
               \theta(x-y)
                 \left(
                 \frac{\lambda}{x} + \frac{h/N_c}{Z_\phi x + M_\phi^2}
                 \right)
                + \theta(y-x)
                 \left(
                   \frac{\lambda}{y}+\frac{h/N_c}{Z_\phi y + M_\phi^2}
                 \right)
             \right]
             \frac{yB(y)}{y +B(y)^2} \;.
             \nonumber\\ \nopagebreak[4]
       \label{eq:SDEQ}
\end{eqnarray}
This technical simplification does not change the qualitative
structure of the SD equation \cite{kn:Kond92}.
\par

Eq.\Ref{eq:SDEQ} is readily converted into
a set of differential equations;
\begin{Eqnarray}
  \frac{d}{dx}B(x)
   &=& -\left( \frac{\lambda}{x}
             + \frac{h}{N_c} \frac{Z_\phi x}{(Z_\phi x+ M_\phi^2)^2}
        \right) V(x),
   \label{eq:SDeqB1}\\
   \frac{d}{dx} \left(xV(x)\right)
   &=& \frac{xB}{x+B(x)^2}
\label{eq:SDeqB2}
\end{Eqnarray}
plus  infrared(IR) and UV  boundary conditions (BC);
\begin{Eqnarray}
    V(0) &=& 0
   \hspace{8em}
    \mbox{(IRBC)}
    \label{eq:IRBC} \\
    \nonumber\\
    \DISP
   \left(
     4 {\Lambda^2 \over M_\phi^2}
      + \frac{1}{N_c} \frac{\Lambda^2}{Z_\phi \Lambda^2 + M_\phi^2}
   \right) h
   &=&
        \left( \frac{B(\Lambda^2)}{V(\Lambda^2)}-\lambda \right)
    \qquad \mbox{(UVBC)} \label{eq:UVBC}\; ,
\end{Eqnarray}
with the ``condensation function'' $V(x)$ being given by
\begin{equation}
  V(x) =  \frac{1}{x}\int_0^{x} dy\frac{yB(y)}{y+B^2(y)}.
\label{eq:condfunc}
\end{equation}
Eqs.\Ref{eq:SDeqB1}--\Ref{eq:UVBC} are the equations that
our analysis is actually based on.
Note here that the fermion pair condensate is related to $V(x)$ as
\begin{equation}
  \VEV{\bar\psi\psi}=-\frac{N_c}{2\pi^2}\Lambda^2 V(\Lambda^2).
  \label{eq:condfunc2}
\end{equation}

\section{Nambu-Goldstone boson decay constant $F_\pi$}
  As we have mentioned before, we need to calculate physical quantity
other than the dynamical mass of the fermion to determine the
``RG''  flow of the bare parameters.
Though the mass of the physical scalar boson $m^{\rm phys}_\phi$
would be the first candidate for this purpose,
we need to solve the SD equation for the scalar boson propagator
in order to calculate  $m^{\rm phys}_\phi$,
which is far beyond the scope of the present paper.
We  therefore choose the decay constant of the NG boson
$F_\pi$.
\par

  One might suspect that the $F_\pi$ is already calculated as the
VEV of the elementary scalar field $\VEV{\sigma}$
through Eq.\Ref{eq:tadpole}.
This is not true in this model, however, because of the presence of
the mixing of $\pi$ with the composite pseudoscalar field.
On might also think that the Pagels-Stokar (PS) formula\cite{kn:PS79}
 of the (composite)
pion decay constant in QCD is applicable to this problem.
However, it cannot be used as it stands due to the mixing with the
elementary pseudoscalar field.
  Actually, the
wave function of the real NG boson  contains the fermion
composite part $\VEV{0|T\psi(x)\bar\psi(y)|{\rm NG}}$
as well as the elementary pseudoscalar part
$\VEV{0|\Phi(x)|{\rm NG}}$ in the gauged Yukawa model.

In this section, we incorporate the mixing effect into the
PS formula for $F_\pi$ and obtain a generalized expression
for $F_\pi$ (``generalized PS formula'') in such a model.
We find that square of the decay constant $F_\pi^2$
of the diagonalized NG
state  is divided into two parts:  $F_\pi^2=F_b^2+F_f^2$,
with $F_b$ and $F_f$ being the bosonic part and the fermionic part,
respectively. The fermion part is evaluated by a certain integral
formula of the fermion mass function $B(x)$.
This result is actually in accord with the well-known result in the
two (elementary) doublet model: the square of the decay constant
$F_\pi^2$ is expressed by the sum of the $({\rm VEV})^2$ of each
doublet.

The NG boson decay constant $F_\pi$ is defined by
\begin{equation}
  \VEV{0|J_5^{a\mu}(x)|{\rm NG}^b(q)}
  = -i q^\mu F_\pi e^{iq\cdot x}\delta^{ab},
\label{eq:fpidef}
\end{equation}
with $|{\rm NG}^b(q)\rangle$ being the (diagonalized) NG boson state
with momentum $q$ ($q^2=0$) and isospin $b$.
The Noether current $J_5^{a\mu}$ is given by
\begin{equation}
  J_5^{a\mu}
  = -i\frac{N_c Z_\phi}{4}
      \tr( \tau^a \Phi^\dagger \partial^\mu \Phi
         - \tau^a \Phi \partial^\mu \Phi^\dagger )
    - \bar\psi \gamma^\mu \gamma_5 \frac{\tau^a}{2} \psi.
\end{equation}

The calculation of $F_\pi$ becomes straightforward, if we know exactly
the wave function of the NG boson:
\begin{Eqnarray}
  \chi_{P\Phi}^a
    &:=& \VEV{0|\Phi(0)|{\rm NG}^a(q)},
    \\
  S(p)\chi_{P\bar\psi\psi}^a(p,q) S(p-q)
    &:=& \int d^4 x e^{-ip\cdot x}
    \VEV{0|T\psi(x)\bar\psi(0)|{\rm NG}^a(q)},
\end{Eqnarray}
where  $S(p)$ is  the fermion propagator and $q^2=0$.
The decay constant $F_\pi$ may be expressed in terms of the wave
function as (see Fig.\ref{fig:Fpi}):
\begin{eqnarray}
  F_\pi q^\mu \delta^{ab}
  &=& \frac{N_c Z_\phi}{4} iq^\mu
      \tr(\{\tau^a, \VEV{\Phi^\dagger}\} \chi_{P\Phi}^b)
  \nonumber\\
  & & -\int\frac{d^4 p}{(2\pi)^4 i}
      \tr(\gamma^\mu \gamma_5\frac{\tau^a}{2} S(p)
           \chi_{P\bar\psi\psi}^b(p,q) S(p-q)).
\label{eq:fpiform}
\end{eqnarray}
Generally speaking, however, it is quite difficult problem to solve
the wave function of composite particles (Bethe-Salpeter amplitude).
Fortunately, this problem is simplified for the NG boson state,
thanks to the Ward-Takahashi identities of the broken symmetry:
\begin{Eqnarray}
\lefteqn{
  \partial_\mu^x \VEV{T J_5^{a\mu}(x) \Phi(y)}
} \nonumber\\
    &=& \{ \frac{\tau^a}{2}, \VEV{\Phi(y)}\} \delta^{(4)}(x-y),
\label{eq:WT1}
    \\
\lefteqn{
  \partial_\mu^x \VEV{T J_5^{a\mu}(x) \psi(y) \bar\psi(z)}
} \nonumber\\
    &=& \frac{\tau^a}{2} \gamma_5 \VEV{T \psi(y) \bar\psi(z)}
    \delta^{(4)}(x-y)
   +\VEV{T \psi(y) \bar\psi(z)}\frac{\tau^a}{2} \gamma_5
              \delta^{(4)}(x-z).
\label{eq:WT2}
\end{Eqnarray}
The spectral representation of (\ref{eq:WT1}) and
(\ref{eq:WT2}) is written as
\begin{Eqnarray}
\lefteqn{
  \sum_n \VEV{0|\Phi(0)|n(q)}
    \frac{q_\mu}{q^2-m_n^2}
    \VEV{n(q)|J_5^{a\mu}(0)|0}
} \nonumber \\
  &=& \{\frac{\tau^a}{2}, \VEV{\Phi}\},
  \\
\lefteqn{
  \int d^4 y e^{-ip\cdot y}
  \sum_n \VEV{0|T\psi(y)\bar\psi(0)|n(q)}
    \frac{q_\mu}{q^2-m_n^2}
    \VEV{n(q)|J_5^{a\mu}(0)|0}
} \nonumber\\
  &=&  S(p)\left( S^{-1}(p)\frac{\tau^a}{2}\gamma_5
               -\frac{\tau^a}{2}\gamma_5 S^{-1}(p-q) \right) S(p-q).
\end{Eqnarray}
Among $| n^a(q)\rangle$ the only
state which survives in $q_\mu=0$ limit is the
 NG boson state $|{\rm NG}^a(q)\rangle$ ($m_{\rm NG}^2=0$).
We therefore obtain the wave functions $\chi_{P\Phi}^a$,
$\chi_{P\bar\psi\psi}^a(p,q=0)$ which are completely determined
at zero-momentum $q=0$ by the
VEV of scalar field and the fermion propagator, respectively:
\begin{Eqnarray}
  \chi_{P\Phi}^a
    &=& -\frac{i}{F_\pi} \{ \frac{\tau^a}{2},\VEV{\Phi} \},
\label{eq:fpiwave1}
    \\
  \chi_{P\bar\psi\psi}^a(p,q=0)
    &=& -\frac{i}{F_\pi} \{ \frac{\tau^a}{2}\gamma_5, S^{-1}(p) \}.
\label{eq:fpiwave2}
\end{Eqnarray}

Following the paper of Pagels and Stokar\cite{kn:PS79,kn:Tana92},
we approximate the fermionic wave function
\begin{equation}
  \chi_{P\bar\psi\psi}^a(p,q) \simeq \chi_{P\bar\psi\psi}^a(p,q=0).
\end{equation}
Now, it is straightforward to calculate the decay constant $F_\pi$ by
plugging the wave function
Eqs.(\ref{eq:fpiwave1})--(\ref{eq:fpiwave2})
into Eq.(\ref{eq:fpiform}).
We find
\begin{Eqnarray}
  F_\pi^2 &=& F_b^2 + F_f^2,
\label{eq:Fpi} \\
  F_b^2 &=& N_c
   Z_\phi \VEV{\sigma}^2,
\label{eq:Fpib} \\
  F_f^2 &=& \frac{N_c}{4\pi^2} \int_0^{\Lambda^2} \!\! dx x
          \frac{B^2(x)-\frac{x}{4}\frac{d}{dx}B^2(x)}
               {(x+B^2(x))^2},
\label{eq:Fpif}
\end{Eqnarray}
where $F^2_b$ comes from the elementary scalar wave function
Eq.(\ref{eq:fpiwave1})
and $F^2_f$ from the fermion composite wave function
Eq.(\ref{eq:fpiwave2}).

Remarkably enough, this generalized PS formula definitely predicts
that the fermion mass gets lower
in the gauged Yukawa model ($Z_\phi > 0$) than in the gauged NJL
model ($Z_\phi =0$) for the same $F_\pi$. Actually it dictates
$
 F_f^2 = F_\pi^2 - F_b^2 < F_\pi^2
$
 for
$
F_b^2  = N_c Z_\phi \VEV{\sigma}^2 > 0.
$
Now, the fermion mass $M^2$ is determined by the formula for $F_f^2$,
Eq.\Ref{eq:Fpif}, as an increasing function of $F_f^2$,
which was the essence of the top quark mass determination
in the top quark condensate model \cite{kn:MTY89,kn:Yama92}.
Thus
we have $M^2\hbox{(gauged Yukawa)} < M^2\hbox{(gauged NJL)}$, since
$F_f^2 \hbox{(gauged Yukawa)} <
 F_\pi^2 = F_f^2 \hbox{(gauged NJL)}$.
\section{Numerical study}
Having set up
the machineries Eqs.\Ref{eq:SDeqB1}--\Ref{eq:UVBC} and
Eqs.\Ref{eq:Fpi}--\Ref{eq:Fpif}, we are now ready to perform
numerical analysis of the phase and ``RG'' structure in
the gauged Yukawa model.
As we mentioned before, we take $M_\phi^2>0$ so that the
chiral symmetry breaking can only be caused by the fermion dynamics.
Because of the redundancy of the parameters in the Lagrangian
Eq.(\ref{eq:Lagrangian}),
we may fix one parameter $M_\phi=\Lambda$ for
positive $M_\phi^2$ without loss of generality.

The goal of the numerical analysis here is to determine the bare
parameters $(h,\lambda,Z_\phi)$ as functions of the ultraviolet cutoff
$\Lambda$ and the ``physical quantities'' $(M, F_\pi,\lambda_R)$;
\begin{equation}
 h(\Lambda,M,F_\pi,\lambda_R), \qquad
 \lambda(\Lambda,M,F_\pi,\lambda_R), \qquad
 Z_\phi(\Lambda,M,F_\pi,\lambda_R),
\label{eq:barepara}
\end{equation}
with $M :=B(0)$ being the dynamical mass of the fermion,
and $\lambda_R$ the renormalized gauge coupling.
Dimensional analysis of Eq.(\ref{eq:barepara}) shows that
$
  h(\Lambda,M,F_\pi,\lambda_R)=h(\Lambda/M,1,F_\pi/M,\lambda_R)
$ and
$
  Z_\phi(\Lambda,M,F_\pi,\lambda_R)
  =Z_\phi(\Lambda/M,1,F_\pi/M,\lambda_R)
$.
Thus $M$ can be set unity, $M=1$, without loss of generality.
Since we restrict ourselves to the standing gauge coupling,
\footnote{
    It is rather straightforward to incorporate running effect of the
   gauge coupling as was done in the gauged NJL model \cite{kn:KSY91}.
}
we do not consider the RG flow to the gauge coupling
$\lambda$ direction: $\lambda_R=\lambda$.
We also disregard the RG flow to the $\lambda_\phi$
direction,
since feedback of running of $\lambda_\phi$ is expected to be small
as we already argued in Section 2.
We thus obtain the ``RG'' flows in $(h,\lambda,Z_\phi)$
space (more precisely,
$(h,Z_\phi)$ plane sliced for each $\lambda$) as we vary $\Lambda$
while keeping $M$ and $F_\pi$ fixed in Eq.(\ref{eq:barepara}).
\par
Numerical calculations are performed as follows in our program.
For given gauge coupling $\lambda$, we solve the differential
equations Eqs.\Ref{eq:SDeqB1}--\Ref{eq:SDeqB2};
\begin{eqnarray}
{d \over dx} B(x) &=& - \left( {\lambda \over x}
+ \frac{\hat h}{N_c}
   {Z_\phi x \over (\hat Z_\phi x+\Lambda^2)^2} \right) V(x),
\nonumber\\
{d \over dx} V(x) &=& - {V(x) \over x} + {B(x) \over x+B^2(x)},
\end{eqnarray}
together with the differential form of
$F_f^2=F^2_f(\Lambda^2)$ Eq.\Ref{eq:Fpif};
\begin{equation}
{d \over dx} F_f^2(x) =
\frac{N_c}{4\pi^2}{x B(x) [B(x)-{x \over 2}B'(x)] \over (x+B^2(x))^2}
\end{equation}
plus the respective IRBC's;
\begin{equation}
  B(0) = 1, \qquad
  V(0) = 0, \qquad
  F_{f}^2(0) = 0,
\end{equation}
with $(\hat h,\hat Z_\phi)$ being trial parameters.
Then we obtain the next trial parameters $(\hat h', \hat Z_\phi')$
from Eqs.(\ref{eq:UVBC}),(\ref{eq:Fpi})--\Ref{eq:Fpif}:
\begin{eqnarray}
\hat h'
  &=& \left(4 + \frac{1}{N_c}\frac{1}{\hat Z_\phi+1} \right)^{-1}
      \left({B(\Lambda^2) \over V(\Lambda^2)} - \lambda \right),
  \nonumber\\
\hat Z_\phi'
  &=& \frac{\pi^2}{2 N_c}
  {F_\pi^2 -F_f^2(\Lambda^2) \over \hat h V^2(\Lambda^2)} \;.
\end{eqnarray}
We repeat the above steps after substituting the trial parameters
$(\hat h', \hat Z_\phi')$ into $(\hat h, \hat Z_\phi)$, and so on.
After several iterations, we obtain the function
Eq.(\ref{eq:barepara}) with sufficient accuracy.
\par

Let us now turn to the result of our numerical calculations.
We here present numerical result only for the $U(1)$ case, i.e.,
$C_F = 1$ and $N_c = 1$ in Eqs.\Ref{eq:couplings},
\Ref{eq:Fpib}--\Ref{eq:Fpif}. Actually,
 the $N_c$ dependence is not significant for the essential
feature of the phase diagram as we will demonstrate by the
analytical study
in the next section.
When we discuss the $N_c>1$ effect, we also have to
include the running effects of the gauge coupling. Numerical study
including both effects will be given in a separate paper.
\par

We first discuss $\lambda=0$ case (pure Yukawa model with
vanishing gauge coupling).
In Fig. \ref{fig:yukawa} we show the critical line
 in the space $(h, Z_\phi)$,
  which separates the chiral symmetric phase $(M =0)$ at
$h < h_c$ and the spontaneously broken phase $(M \ne 0)$ at
$h > h_c$, with $h_c = h_c(Z_\phi)$ being a critical
coupling for each $Z_\phi$.\footnote{
In the previous work\cite{kn:KTY90} the case of $Z_\phi=1$
was investigated by
varying the scalar mass $M_\phi$.  It was shown (see Fig. 3 of
Ref.\cite{kn:KTY90}) that there exists a critical
Yukawa coupling $h_c$ and the dependence of $h_c$ on $M_\phi$ is
very small.
}
   This is confirmed by the
analytical study in the next section.
Above the critical coupling $h > h_c$, the chiral condensate
$\VEV{\bar \psi \psi}$ exhibits a characteristic behavior shown in
Fig. \ref{fig:orderparameter}, which was first obtained
 in Ref.\cite{kn:KTY90}.
Note that the critical Yukawa coupling $h_c$
 monotonically increases in
$Z_\phi$ and the dependence of $h_c$ on $Z_\phi$
is very small.
This also agrees with the analytical study.
\par

Next we consider another limit, $Z_\phi=0$, which corresponds to the
gauged NJL model.  Actually, lines of the equi-correlation length
 $\xi_f = \Lambda/M$ in the $Z_\phi=0$ plane
are depicted in Fig. \ref{fig:gNJL} for $N_c=1$, which converge
in the $\xi_f \rightarrow \infty$ limit toward the critical line
(bold line in Fig. \ref{fig:gNJL}). This critical line is
 nothing but the well-known critical line\cite{kn:KMY89} of the
 gauged NJL model written in the space $(h,\lambda)$:
\begin{equation}
 \left(4 + \frac{1}{N_c} \right) h
  =\frac14 (1+\sqrt{1-\lambda/\lambda_c})^2
  \label{eq:gNJLcline}
\end{equation}
for $\lambda<\lambda_c=1/4$. (This analytical
 expression will be derived in the next section.)
\par

Then in Fig.\ref{fig:length}
the surface of the equi-$\xi_f$ is depicted
in the three-dimensional bare parameter space ($h, \lambda, Z_\phi$)
 for the
Yukawa coupling $h$, the gauge coupling $\lambda$
and the ``hopping parameter'' of the scalar field $Z_\phi$.
The $\xi_f \rightarrow \infty$ limit is the critical surface
$h=h_c(\lambda, Z_\phi)$
separating the chiral symmetric phase ($h<h_c$)
and the spontaneous symmetry
breaking  phase ($h>h_c$), which is shown by the surface with mark
``$\times$'' in
Fig. \ref{fig:length}. Below the critical surface ($h<h_c$),
there is no
nontrivial solution for the SD equation($M=0$).
\par

Now to the RG flow, which is obtained by
fixing the NG boson decay constant $F_\pi$ as well as $M$.
Given the value of $F_\pi$, we
obtain the $F_\pi$-constant surface (two-dimensional manifold)
in the space
($h, \lambda, Z_\phi$), which is then sliced
by the $\lambda$-fixed plane
in accord with our setting of the standing gauge coupling.
RG flows are obtained as the
intersection of the $F_\pi$-constant surface
with the $\lambda$-fixed planes,
see Fig. \ref{fig:flow}.
 Imagine the situation that each RG flow passes
across the equi-$\xi_f$ surfaces of
Fig. \ref{fig:length}.
  Then each flow curve
may be considered to be parameterized with respect to the value of
$\xi_f$.  We may interpret that  the crossed points of
 a RG flow with the equi-$\xi_f$ surfaces correspond to the result of
 successive steps of the RG transformation.%
\footnote{
  Our analysis based on the SD equation can
  determine the
flow only in the spontaneous chiral symmetry breaking phase where the
  nontrivial solution ($M \ne 0, F_\pi \ne 0$) exists.
  Thus the renormalized
  trajectory cannot be drawn as extended to
  the symmetric phase, which of
  course is an ``artifact'' of this framework.
  We may improve this situation
  by calculating effective potential (or $m^{\rm phys}_\phi$)
  as we did in the gauged NJL model \cite{kn:KTY93a}.
}
\par
All the flows in each figure of
Fig. \ref{fig:flow} are for the same $F_\pi$ and for different
values of $\lambda$. Most of the flows either run to
$Z_\phi \to \infty$
or $Z_\phi \to 0$
for $\xi_f := \Lambda/M \gg 1$,
 but there is an indication of a single
flow (``renormalized trajectory'') terminating at a finite value
of $Z_\phi$ and $\lambda$ (``(nontrivial) fixed point'')
on the critical surface.
We can draw a similar figure
for a different $F_\pi$, which indicates a
different fixed point and the renormalized
trajectory (compare (a) and (b) in
Fig. \ref{fig:flow}). A set of such fixed points form a line
(``fixed line'') which is depicted by the dotted dashed line in
Fig. \ref{fig:flow}. (The precise location of the fixed line is
determined by the analytical study in the next section
where we find that
the fixed line extends from
$(h, \lambda, Z_\phi)$ = ($-1/20$,1/4,0)  to
 (1/4,0,$\infty$) for $N_c=1$.)

We may take another look at this phase diagram by
Fig. \ref{fig:yukawa} ($\lambda =0$)
and Fig. \ref{fig:2dimflow} ($\lambda > 0$), where the flows
are depicted for different values of $F_\pi$
in the fixed-$\lambda$ plane
 $(h,Z_\phi)$. In Fig.\ref{fig:2dimflow} ($\lambda >  0$)
 there is again an
 indication that a single flow (renormalized
 trajectory) terminates at
 a finite $Z_\phi$ (fixed point) on the critical surface,
 while all others run to either $Z_\phi \to \infty$  or
 $Z_\phi \to 0$
 for  $\xi_f \gg 1$.
This is sharply contrasted with the pure
Yukawa model ($\lambda=0$) whose
RG flows are shown in Fig.\ref{fig:yukawa} (solid lines).
There is no indication of the existence of the
``(nontrivial) fixed point''.
This is confirmed by the analytical study in the next section.
This also
 agrees quite well with the result
of the lattice Yukawa model (see Section 6).
 Thus we found that a fixed line exists in the gauged Yukawa model
 solely due to the presence of the gauge coupling.
 This is our main result.
\par

   The RG flow shows that there is
a  ``critical value'' for the decay constant $F_\pi$
which separates the three-dimensional space
(above the critical surface $h>h_c$) into two regions:
\begin{description}
\item[(I)] $F_\pi < F_f^\infty$ $( F_f^\infty :=F_f(\Lambda=\infty)$);
           RG flows extend to $Z_\phi =0$ for  $\xi_f \gg 1$,
\item[(II)] $F_\pi > F_f^\infty$ ;
            RG flows run away to  $Z_\phi \to \infty$ .
\end{description}
The surface $F_\pi = F^\infty_f$
is the boundary  of the region (I) and (II)
which consists of the renormalized trajectories.
The intersection of the critical
surface and the renormalized trajectories forms the fixed line.
\par
Finally, we consider the ratio $F^2_b/F^2_f$ for the NG boson decay
constant $F^2_\pi$, where $F^2_b$ and $F^2_f$ are the bosonic part
and the
fermionic part of $F^2_\pi$,
respectively, as defined by Eqs.\Ref{eq:Fpib}--\Ref{eq:Fpif}.
In Fig. \ref{fig:ratio} the ratio $F^2_b/F^2_f$ is shown
along the RG flow
(solid line) in the fixed-$\lambda$ plane and the broken
line passing through
the origin denotes the fixed-cutoff $\Lambda$ line where
the line with the
steeper slope corresponds to the smaller cutoff.
\par
In the presence of gauge coupling, $F_f$ is finite even
in the infinite cutoff limit ($F_f^\infty<\infty$),
 because the solution
$B(-p^2)$ is damping in the asymptotic UV region
(see \linebreak[3] Eqs.\Ref{eq:asyformZs2},\Ref{eq:asyformZl2}).
The RG flow in the region (II) runs
away to $Z_\phi \rightarrow \infty$ as $\Lambda \rightarrow \infty$.
However, even in the $Z_\phi \rightarrow \infty$ limit,
the ratio
\begin{equation}
{F_b^2 \over F_f^2}  =  {F_\pi^2 \over F_f^2} - 1
\end{equation}
takes a finite non-zero value depending
on the initially specified value of
$F_\pi$  for the RG flow.  As $F_\pi$ approaches to $F_f$ from
the above, the ratio decreases and finally becomes zero
at $\Lambda=\infty$ just for
$F_\pi=F_f^\infty$.
In the region (I) including
the $F_\pi=F_f(\Lambda^2)$ surface, therefore, we can always make the
contribution from the bosonic part $F^2_b$ to $F^2_\pi$
arbitrarily small by
choosing large cutoff or small $Z_\phi$.
This implies that the fermionic
part  $F^2_f$ is dominant in $F^2_\pi$ in the small $Z_\phi$ region
for large $\Lambda$, as
was already noted\cite{kn:KTY90}  in the pure Yukawa model
which has actually only the region (I)\cite{kn:KTY90}.
\par
Comparing two points on the fixed-$\Lambda$ line, we find a  larger
$F^2_\pi$ by increasing $Z_\phi$.
This implies that the contribution of the
fermion dynamical mass $M$ to $F_\pi$ gets smaller
by increasing $Z_\phi$,
since we fixed the fermion mass $M=B(0)=1$
in our numerical calculation.
Here we notice again that the gauged Yukawa model reduces
to the gauged NJL
model in the limit $Z_\phi=0$. Therefore this result
can be rephrased as
follows; by including the Yukawa interaction
due to non-zero $Z_\phi$, the
fermion dynamical mass $M$
can be reduced for a fixed value of $F_\pi$.
This is actually a direct consequence of the salient feature of
our generalized PS formula Eqs.\Ref{eq:Fpi}--\Ref{eq:Fpif},
as we emphasized in the end of Section 3.
This mechanism may be applied to a scenario of the
top quark condensate
(``Yukawa-driven Top Mode Model'')\cite{kn:KTY90,kn:KTY93,kn:CR91}
 to lower the mass of the top quark.

\par

\section{Analytical study}
In the previous section, we have solved the nonlinear SD equation
Eq.(\ref{eq:SDEQ}) numerically.
We solve here the SD equation Eq.(\ref{eq:SDEQ})
in a analytical manner
by making use of
the linearization (bifurcation)\cite{kn:Atki87,kn:Kond92},
 which is valid near the critical point ($M \ll \Lambda$).
We thus obtain here an analytical expressions of the critical surface
and the fixed line.  These results actually confirm the numerical
calculations in the previous section.

Let us start with the bifurcation approximation of Eq.(\ref{eq:SDEQ}).
Near the critical point $B(x)\ll\Lambda$, the integrand can be safely
replaced by a linearized expression,
$x B(x)/(x+B(x)^2)\rightarrow B(x)$
for dominant region of the integral $x\gtrsim M^2$ with $M=B(M^2)$.
Small but still important nonlinear effect from the region
$x\lessim M^2$ can be properly taken into account by introducing the
infrared cutoff $M$ in Eq.(\ref{eq:SDEQ})
except for some subtlety to be discussed later in the pure NJL limit.

Under this approximation, the differential equation Eq.\Ref{eq:SDeqB1}
and UVBC Eq.\Ref{eq:UVBC} remain the same,
while the Eq.\Ref{eq:SDeqB2}
and IRBC Eq.\Ref{eq:IRBC} are changed into
\begin{Eqnarray}
  &&
  \frac{d}{dx}\left( x V(x) \right)= B(x) \label{eq:dfeqB} \;,
  \\
  &&
  V(M^2) = 0, \qquad B(M^2)=M \;,\label{eq:IRBC2}
\end{Eqnarray}
respectively.
 From equation Eq.\Ref{eq:dfeqB} and Eq.\Ref{eq:SDeqB1}
we obtain the differential equation for $V(x)$:
\begin{equation}
        \frac{d^2}{dx^2} \left(xV(x)\right) +\left
(        \frac{\lambda}{x} + \frac{h}{N_c} \frac{Z_\phi x}{(Z_\phi x+
M_\phi^2)^2} \right) V(x) = 0 \; . \label{eq:dfeq}
\end{equation}
\par

Before starting the full analysis of Eq.(\ref{eq:SDEQ}) with
bifurcation approximation, we briefly describe the result in the
$1/N_c\rightarrow 0$ limit, in which the analytical calculation is
greatly simplified without changing the qualitative feature.
In this limit, the differential equation Eq.(\ref{eq:dfeq}) and its
boundary conditions are  simplified:
\begin{eqnarray}
 & &
  \frac{d^2}{dx^2} \left(xV(x)\right) +\frac{\lambda}{x} V(x) = 0,
  \qquad
  B(x)=\frac{d}{dx}(xV(x)),
 \nonumber\\
 & &
  V(M^2)=0, \qquad B(M^2)=M, \qquad
  4\frac{\Lambda^2}{M_\phi^2} h
  =  \frac{B(\Lambda^2)}{V(\Lambda^2)}-\lambda.
\label{eq:1/Nc1}
\end{eqnarray}
The solution of Eq.(\ref{eq:1/Nc1}) is given by
\begin{Eqnarray}
  V(x)&=&\frac{M}{2\mu}\left[
           \left(\frac{x}{M^2}\right)^{-1/2+\mu}
          -\left(\frac{x}{M^2}\right)^{-1/2-\mu}
         \right],
      \\
  B(x)&=&\frac{M}{2\mu}\left[
           \left(\frac{1}{2}+\mu\right)
             \left(\frac{x}{M^2}\right)^{-1/2+\mu}
          -\left(\frac{1}{2}-\mu\right)
             \left(\frac{x}{M^2}\right)^{-1/2-\mu}
         \right],
\end{Eqnarray}
where $2\mu:=\sqrt{1-4\lambda}$ and $M/\Lambda\;(=\xi_f^{-1})$
is calculated as
\begin{equation}
  \left(\frac{1}{\xi_f}\right)^{4\mu}
  = \left(\frac{M}{\Lambda}\right)^{4\mu}
  = \frac{4 h -(1/2+\mu)^2 r_\phi}{4 h -(1/2-\mu)^2 r_\phi},
\label{eq:1/Nscaling}
\end{equation}
with $r_\phi:=M_\phi^2/\Lambda^2$.
 From Eq.\Ref{eq:1/Nscaling} we obtain the critical surface
\begin{equation}
 4 h=(1/2+\mu)^2 r_\phi,
\label{eq:NcInfCri}
\end{equation}
which is independent of $Z_\phi$.
Eq.\Ref{eq:NcInfCri} coincides with the critical
line of the gauged NJL model\cite{kn:KMY89}.
\par
We next discuss the RG flow at  the $1/N_c$ leading order.
The function $h(\Lambda, M, F_\pi, \lambda_R=\lambda)$ of
Eq.(\ref{eq:barepara}) can be read from Eq.(\ref{eq:1/Nscaling}):
\begin{equation}
  h(\Lambda, M, F_\pi, \lambda)
  = \frac{r_\phi}{4}
    \frac{(1/2+\mu)^2 - (1/2-\mu)^2 (M/\Lambda)^{4\mu}}
         {1-(M/\Lambda)^{4\mu}}.
\label{eq:scal/N1}
\end{equation}
  From Eq.\Ref{eq:Fpi}--\Ref{eq:Fpib}
the function $Z_\phi(\Lambda, M, F_\pi, \lambda)$ is given by
\begin{equation}
  Z_\phi(\Lambda, M, F_\pi, \lambda)
  = \frac{\pi^2}{2 N_c} r_\phi^2
    \frac{F_\pi^2 - F_f^2(\Lambda^2)}
         {h(\Lambda, M, F_\pi, \lambda) V^2(\Lambda^2)},
\label{eq:scal/N2}
\end{equation}
where the IR cutoff $M^2$ was introduced:
\begin{equation}
  F_f^2 = \frac{N_c}{4\pi^2} \int_{M^2}^{\Lambda^2} \!\! dx x
          \frac{B^2(x)-\frac{x}{4}\frac{d}{dx}B^2(x)}
               {(x+B^2(x))^2},
\end{equation}
since the mass function $B(x)$ is determined only for $x>M^2$ in the
bifurcation approximation.
Among these ``RG'' flows for various $F_\pi$ and $M$, the
``renormalized trajectory'' possesses a distinguished property:
The functions $h$ and $Z_\phi$ remain finite in the limit
$\Lambda/M\rightarrow\infty$ (fixed point):
\begin{Eqnarray}
  h(\Lambda,M,F_\pi,\lambda)
    &=& h(\Lambda/M,1,F_\pi/M,\lambda)
        \rightarrow h(\infty, 1, F_\pi/M, \lambda),
    \\
  Z_\phi(\Lambda,M,F_\pi,\lambda)
    &=& Z_\phi(\Lambda/M,1,F_\pi/M,\lambda)
        \rightarrow Z_\phi(\infty,1,F_\pi/M,\lambda).
\end{Eqnarray}
Noting that $V(\Lambda^2)\rightarrow 0$ for
$\Lambda/M\rightarrow \infty$,
we can  easily see from Eq.(\ref{eq:scal/N1})
and Eq.(\ref{eq:scal/N2})
that this condition can be satisfied only for
\begin{equation}
  F_\pi^2=F_f^2(\Lambda^2=\infty),
\label{eq:traj/N}
\end{equation}
otherwise $Z_\phi\rightarrow \pm \infty$ for
$\Lambda/M\rightarrow\infty$.

The mass function $B(x)$ is independent of $(h,Z_\phi)$
 which depends
on $\Lambda$ through Eq.(\ref{eq:scal/N1}) and Eq.(\ref{eq:scal/N2}).
Thus, it is also independent of the cutoff $\Lambda$ and we can easily
calculate the function
$\bar F_f^2(\Lambda^2):=F_f^2(\infty)-F_f^2(\Lambda^2)$ by
\begin{eqnarray}
  \bar F_f^2(\Lambda^2)
    &=& \frac{N_c}{4\pi^2} \int_{\Lambda^2}^{\infty} \!\! dx x
        \frac{B^2(x)-\frac{x}{4}\frac{d}{dx}B^2(x)}
             {(x+B^2(x))^2}
     \simeq  \frac{N_c}{4\pi^2} \int_{\Lambda^2}^{\infty} \!\! dx
              \frac{B^2(x)-\frac{x}{4}\frac{d}{dx}B^2(x)}{x}
   \nonumber\\
   &\simeq& \frac{N_c}{4\pi^2} \frac{M^2}{4\mu^2}
       \frac{(1/2+\mu)^2(5/4-\mu/2)}{1-2\mu}
       \left(\frac{M}{\Lambda}\right)^{2-4\mu},
\label{eq:barF/N}
\end{eqnarray}
where we have made an approximation
$B(x>\Lambda^2)=(M/2\mu)(1/2+\mu)(x/M^2)^{-1/2+\mu}$.

Plugging Eq.(\ref{eq:traj/N}) and Eq.(\ref{eq:barF/N})
into Eq.(\ref{eq:scal/N1}) and
Eq.(\ref{eq:scal/N2}), we obtain the explicit form of
the RG flow on the renormalized
trajectory for $M \ll \Lambda$ in the $1/N_c \to 0$ limit:
\begin{Eqnarray}
  & &
  h(\Lambda,M,F_f(\infty),\lambda)
  = \frac{r_\phi}{4}(1/2+\mu)^2 \left(
        1 +
        \frac{8\mu}{(1+2\mu)^2}\left(\frac{M}{\Lambda}\right)^{4\mu}
      \right),
\label{eq:equxi1/N}
  \\
  & &
  Z_\phi(\Lambda,M,F_f(\infty),\lambda)
  h(\Lambda,M,F_f(\infty),\lambda)=
  \frac{r_\phi^2}{8} \frac{(5/4-\mu/2)(1/2+\mu)^2}{1-2\mu},
\label{eq:rentrj1/N}
\end{Eqnarray}
where we estimated
$V(\Lambda^2)\simeq(M/2\mu)(M/\Lambda)^{1-2\mu}$.
It should be emphasized that the disappearance of the ``renormalized
trajectory'' at $\lambda=0$ can be already observed at this stage:
$Z_\phi \to \infty$ as $\lambda \to 0$.

Now we return to the analysis of Eq.(\ref{eq:SDEQ}) for finite $N_c$.
   Rewriting Eq.\Ref{eq:dfeq} in terms of a new variable
$\eta:=Z_{\phi}x/M_\phi^2$ for $v(\eta)$:
\begin{equation}
        V(x) = \eta^{-1/2+\mu}(1+\eta)^{1/2+\nu} v(\eta),
\label{eq:Vtov}
\end{equation}
   we obtain a differential equation for $v$:
\begin{equation}
        \eta(1+\eta)v''(\eta)   + ( 1+2\mu +2(1+ \mu+\nu)\eta) v'
+ \frac12 (1+2\mu)(1+2\nu)v =0,         \label{eq:hyp}
\end{equation}
where the prime denotes the differentiation
with respect to $\eta$, $v':=\frac{d}{d\eta}v$,
and $\mu$ and $\nu$ are defined by
\begin{equation}
  \mu :=\frac12\sqrt{1-4\lambda} \;,\quad \nu :=\frac12
\sqrt{1-4h/Z_\phi N_c}\;.
\end{equation}
\par
 The general solution of  Eq.\Ref{eq:hyp} is obtained
as a linear combination of two
hypergeometric functions.
The ratio of the two coefficients of
the linear combination of the hypergeometric functions is
determined from the IRBC Eq.\Ref{eq:IRBC2}.
Then we obtain the solution of Eq.\Ref{eq:dfeq}:
\begin{equation}
        V(\eta) = \kappa \left[ V_{+}(\eta) - V_{-}(\eta) \right],
\end{equation}
where $\kappa$ is a normalization constant
and $V_{\pm}(\eta)$ is defined as
\begin{equation}
        V_{\pm}(\eta) :=
        \left(\frac{\eta}{\eta_M}\right)^{-1/2 \pm \mu}
        \left( \frac{1+\eta}{1+\eta_M} \right)^{1/2 \pm \nu}
      \frac{ F(\frac12\pm\mu\pm\nu+\omega,\frac12\pm\mu\pm\nu-\omega,
           1\pm 2\mu; -\eta)}
         { F(\frac12\pm\mu\pm\nu+\omega,\frac12\pm\mu\pm\nu-\omega,
           1\pm 2\mu; -\eta_M)},
\end{equation}
with $\eta_M :=Z_\phi M^2/M_\phi^2$
and
\begin{equation}
   \omega :=\frac12 \sqrt{1-4\lambda-4h/Z_\phi N_c}\;.
\end{equation}
Then the mass function $B(x)$ is obtained from Eq.\Ref{eq:dfeqB}
as follows:
\begin{eqnarray}
        B(x) &=& \frac{d}{d\eta}( \eta V(\eta)) \nonumber\\
        &=&
        \kappa  \Biggl[
          \left(
                 \frac12 +\mu
                 + (\frac12+\nu) \frac{\eta}{1+\eta}
                 - \tilde{V}_{+}(\eta) )
          \right) V_{+}(\eta)
          \nonumber\\
          &&-
          \left(
                 \frac12 -\mu
                 + (\frac12-\nu) \frac{\eta}{1+\eta}
                 - \tilde{V}_{-}(\eta)
          \right) V_{-}(\eta)
          \Biggr] ,
\end{eqnarray}
where $\tilde{V}_{\pm}$ are defined by
\begin{equation}
        \tilde{V}_{\pm}(\eta) := (1/2\pm\nu)\eta
          \frac{ F(\frac32\pm\mu\pm\nu+\omega,
                    \frac32\pm\mu\pm\nu-\omega,
                       2\pm 2\mu; -\eta)}
               { F(\frac12\pm\mu\pm\nu+\omega,
                   \frac12\pm\mu\pm\nu-\omega,
                       1\pm 2\mu; -\eta)}
                \mbox{ .}
\end{equation}
The coefficient $\kappa$ is determined by
the normalization condition $B(M^2)= M$.
\par
It is worth remarking that the asymptotic form of the solution
for $V(x)$ and $B(x)$ are given as follows:
\begin{Eqnarray}
    V(\eta) &=&  C_1 \eta^{-{1 \over 2}+\mu}
                + C_2 \eta^{-{1 \over 2}-\mu},
     \label{eq:asyformZs1}\\
    B(x)  &=&  C_1 ({1 \over 2}+\mu) \eta^{-{1 \over 2}+\mu}
             + C_2 ({1 \over 2}-\mu) \eta^{-{1 \over 2}-\mu}
\label{eq:asyformZs2}
\end{Eqnarray}
in the small $Z_\phi$ region, and
\begin{Eqnarray}
    V(\eta) &=&  C_1' \eta^{-{1 \over 2}+\omega}
                + C_2' \eta^{-{1 \over 2}-\omega},
     \label{eq:asyformZl1}\\
    B(x)  &=&  C_1' ({1 \over 2}+\omega) \eta^{-{1 \over 2}+\omega}
             + C_2' ({1 \over 2}-\omega) \eta^{-{1 \over 2}-\omega}
\label{eq:asyformZl2}
\end{Eqnarray}
in the large $Z_\phi$ region,
where $C_1$, $C_2$, $C_1'$ and $C_2'$ are constants determined
 by the IRBC Eq.\Ref{eq:IRBC2} and the normalization condition
 $B(M^2)=M$.
\par
    From the UVBC Eq.\Ref{eq:UVBC}
 we obtain the scaling relation, equi-$\xi_f\;(:= \Lambda/M)$ surface,
 of the fermion dynamical mass for
     three parameters $(h, \lambda, Z_\phi)$:
\begin{eqnarray}
        \lefteqn{
            \left( \frac{M}{\Lambda} \right)^{4\mu}
            \left(
         \frac{Z_\phi \frac{M^2}{\Lambda^2} + r_\phi}{Z_\phi +r_\phi}
            \right)^{2\nu}
           P( {Z_\phi \over r_\phi} \frac{M^2}{\Lambda^2})
          }
        \nonumber\\
        &&=
        \frac
         {
            (4h/r_\phi-(1/2+\mu)^2)
            - \frac{Z_\phi}{(r_\phi+Z_\phi)}(1/2+\nu)^2
            +  \tilde{V}_{+}(Z_\phi/r_\phi)
         }
         {
            (4h/r_\phi-(1/2-\mu)^2)
            - \frac{Z_\phi}{(r_\phi+Z_\phi)}(1/2-\nu)^2
            + \tilde{V}_{-}(Z_\phi/r_\phi)
         }
         P({Z_\phi \over r_\phi}) \; ,
         \label{eq:scaling}
\end{eqnarray}
where we have defined $r_\phi:= M_\phi^2/\Lambda^2$ and
\begin{equation}
        P(y)   :=
       \frac{ F(\frac12+\mu+\nu+\omega,\frac12+\mu+\nu-\omega,
                1+2\mu;-y)}
            { F(\frac12-\mu-\nu+\omega,\frac12-\mu-\nu-\omega,
                1-2\mu;-y)}
        \; .
\end{equation}
\par
   When we take the limit $Z_\phi \to 0$,
we obtain the scaling relation for the gauged NJL
model\cite{kn:KMY89}:
\begin{equation}
        \left( \frac{M}{\Lambda} \right)^{4\mu}
        = \frac{ (4+1/N_c)h - r_\phi(1/2+\mu)^2}
               {(4+1/N_c)h-r_\phi(1/2-\mu)^2}\;,
\label{eq:gNJLlimit}
\end{equation}
since $P(0)=1$ and $\tilde{V}_{\pm}(0)=0$.
This implies the critical line:
\begin{equation}
 h = h_c(\lambda) = {r_\phi(1/2+\mu)^2 \over 4+1/N_c},
\label{eq:gNJL}
\end{equation}
which yields the critical point of the pure NJL model in the limit
$\lambda \rightarrow 0$:
\begin{equation}
  h_c(0) = {r_\phi \over 4+1/N_c}.
  \label{eq:NJL}
\end{equation}
The critical exponent $\nu_M$ for the fermion mass defined by
\begin{equation}
  {M \over \Lambda} = (h-h_c(\lambda))^{\nu_M}
\end{equation}
takes the continuously changing value (for $\lambda<1/4$):
\begin{equation}
  \nu_M(\lambda) = {1 \over 4\mu} = {1 \over 2\sqrt{1-4\lambda}}.
\end{equation}
For further results on other critical exponents, see, e.g.,
Ref.\cite{kn:Kond92}.

\par
In the vanishing gauge coupling limit,  $\lambda \to 0$,
we obtain the scaling relations for the pure Yukawa model:
\begin{equation}
      \frac{Z_\phi \frac{M^2}{\Lambda^2}+r_\phi}{Z_\phi+r_\phi}
   =
\cases{
    \left[\displaystyle
      \frac{(4r_\phi+4Z_\phi+1/N_c)h - (1/2+\nu) Z_\phi}
           {(4r_\phi+4Z_\phi+1/N_c)h - (1/2-\nu) Z_\phi}
   \right]^{1/2\nu}
   \cr\cr
    \qquad =   \exp \displaystyle
        \left[- {2 \tanh^{-1}
           \frac{ {1 \over 2}Z_\phi \sqrt{1-\frac{4h}{Z_\phi N_c}}}
                {(4Z_\phi+4r_\phi+1/N_c)h - {1 \over 2}Z_\phi}
                  \over \sqrt{1-\frac{4h}{Z_\phi N_c}}}
        \right]
   &($Z_\phi > 4h/N_c$)  \cr \cr
   \exp \left[ \displaystyle - \frac{4}{4Z_\phi+4r_\phi-1/N_c}\right]
   &($Z_\phi = 4h/N_c$)  \cr \cr
   \exp\left[\displaystyle
        - \frac{2n \pi + 2 \tan^{-1}
           \frac{ {1 \over 2}Z_\phi \sqrt{\frac{4h}{Z_\phi N_c}-1}}
                {(4Z_\phi+4r_\phi+1/N_c)h - {1 \over 2}Z_\phi}}
           {\sqrt{\frac{4h}{Z_\phi N_c}-1}}
  \right]
  & ($ 0< Z_\phi < 4h/N_c$)
 \cr }
   \label{eq:yukawalimit}
\end{equation}
where $n$ is an integer ($n=0,1,2,...$).
This scaling relation is not
the same as the previous result\cite{kn:KTY90}
on the pure Yukawa model which does not include the tadpole
     contribution.
     Without tadpole we have the solution only for $Z_\phi < 4h/N_c$.
\par
  In the $Z_\phi \rightarrow \infty$ limit, Eq.\Ref{eq:yukawalimit}
  reduces to
\begin{equation}
 {M^2 \over \Lambda^2} = {4h-1 \over 4h}
  = \exp \left[ -2 \tanh^{-1} \left({1 \over 1+8(h-1/4)}\right)
         \right],
\end{equation}
which implies the critical point $h_c=1/4$.
On the other hand, the $Z_\phi \rightarrow 0$ limit reproduces
Eq.\Ref{eq:NJL} for $n=0$.
Hence the critical line $h=\tilde h_c(Z_\phi)$ extends from
$(h,\lambda,Z_\phi) = (r_\phi/(4+1/N_c),0,0)$ to
$(1/4, 0, \infty)$ in the $\lambda=0$ plane.
\par
   It is worth remarking that the scaling relations
   Eq.\Ref{eq:gNJLlimit} and
Eq.\Ref{eq:yukawalimit} do not correctly reproduces
the scaling relation of
the pure NJL model by taking both limits
$Z_\phi \rightarrow 0$ and $\lambda
\rightarrow 0$ \cite{kn:NSY89},
 although the gauged Yukawa model reduces to the pure NJL
 model in this limit.
  This is because the bifurcation technique cannot be applied to
the pure NJL model in this limit where the mass function has no
$p^2$ dependence.
Note that the above scaling relations obtained in the
bifurcation technique give a meaningful result only
in the neighborhood of the
critical point $M/\Lambda=0$ which is not sensitive to the
nonlinearity of the SD equation.
Therefore the $Z_\phi \rightarrow 0$ limit
of the scaling Eq.\Ref{eq:gNJLlimit}
and the $\lambda  \rightarrow 0$ limit
of Eq.\Ref{eq:yukawalimit} cannot give the correct
scaling relation, apart from the critical point
and the critical exponent.\footnote{
In the pure Yukawa model the critical exponent $\tilde \nu_M$
defined by
$
  {M \over \Lambda} = (h-\tilde h_c(Z_\phi))^{\tilde \nu_M}
$
changes continuously depending on the value  $Z_\phi$
in the case $r_\phi \rightarrow 0$,  while  $\tilde \nu_M$
takes, irrespective of $Z_\phi$, the mean-field value:
$
  \tilde \nu_M = {1 \over 2},
$
as long as $r_\phi\not=0$ in any cutoff $\Lambda$, as was pointed out
in Ref.\cite{kn:Kond89}.
In the latter case, all the critical exponents take their mean field
values.  Actually the chiral condensate takes the critical exponent
 $1/2$,
with which the numerical calculation shown in Fig.2 is consistent.
}
\par
   Now let us consider the $\xi_f :=\Lambda/M \to \infty$
limit of the scaling relation
Eq.\Ref{eq:scaling}, namely, the critical surface
$h = h_c(\lambda,Z_\phi)$
in three-dimensional parameter space $(h, \lambda, Z_\phi)$.
Since the left-hand side of Eq.\Ref{eq:scaling}
vanishes in the limit for $0 < \lambda< \lambda_{c}=1/4$ ($\mu>0$),
the critical surface obeys the equation
\begin{eqnarray}
        [ 4h/r_\phi-(1/2+\mu)^2
          - \frac{Z_\phi}{r_\phi+Z_\phi}(1/2+\nu)^2
            + \tilde{V}_{+}(Z_\phi/r_\phi)]
         P(Z_\phi/r_\phi)  =  0,
         \label{eq:critical}
\end{eqnarray}
which implies
\begin{eqnarray}
  \lefteqn
  {
   \Biggl[
     4h/r_\phi-(\frac12+\mu)^2
     - \frac{Z_\phi}{r_\phi+Z_\phi}(\frac12+\nu)^2
  } \nonumber\\
    &&+ (\frac12+\nu) {Z_\phi \over r_\phi}
    \frac
      { F(\frac32+\mu+\nu+\omega,\frac32+\mu+\nu-\omega,2+2\mu;
           -Z_\phi/r_\phi)}
     { F(\frac12+\mu+\nu+\omega,\frac12+\mu+\nu-\omega,1+2\mu;
           -Z_\phi/r_\phi)}
   \Biggr] \nonumber\\
   &&\times
   \frac
   { F(\frac12+\mu+\nu+\omega,\frac12+\mu+\nu-\omega,1+2\mu;
               -Z_\phi/r_\phi)}
   { F(\frac12-\mu-\nu+\omega,\frac12-\mu-\nu-\omega,1-2\mu;
               -Z_\phi/r_\phi)}
 =  0.
         \label{eq:critical2}
\end{eqnarray}
This is the exact form of the critical surface of
the gauged Yukawa model.
By solving this equation, we obtain the following critical surface:
\begin{eqnarray}
   h = h_c(\lambda,Z_\phi) = r_\phi(1/2 +\mu)^2
      \frac{1+Z_\phi/r_\phi}
           { 4+1/N_c +4Z_\phi/r_\phi
              + \frac{1/N_c}{2+2\mu}\frac{Z_\phi}{r_\phi+Z_\phi}
           }
      \qquad\mbox{for $Z_\phi \ll 1$}\;.
\end{eqnarray}
  The above equation shows the deviation from the critical
line Eq.\Ref{eq:gNJL}
of the gauged NJL model ($Z_\phi=0$) due to the presence of $Z_\phi$.

\par
  Next we consider the decay constant $F_\pi$.
Using Eq.\Ref{eq:tadpole} and Eq.\Ref{eq:condfunc2}
with the IR cutoff $M$,
we may rewrite  Eq.\Ref{eq:Fpi}--\Ref{eq:Fpif} as
\begin{equation}
   Z_\phi = r_\phi^2 \frac{\pi^2}{2N_c}
            \frac{F_\pi^2 - \bar{F}_f^{2}(M^2)}
                 {hV^2(\Lambda^2)}
            + r_\phi^2 \frac{\pi^2}{2N_c}
            \frac{  \bar{F}_f^{2}(\Lambda^2) }
                 {  hV^2(\Lambda^2) } \;,
             \label{eq:Zphi}
\end{equation}
with the function $\bar{F}_f(x)$ being defined by
\begin{equation}
   \bar{F}_{f}^2(x) :=
   \frac{N_c}{4\pi^2}
   \int^{\infty}_{x}dy y
        \frac{ B(y)( B(y)-\frac{y}{2}\frac{d}{dy}B(y))}
             {(y + B^2(y))^2}\;,
\end{equation}
which is finite (for $0 < \lambda < \lambda_c=1/4$)
from the asymptotic form Eqs.\Ref{eq:asyformZs2},\Ref{eq:asyformZl2}
 of the solution $B(x)$.
 $\bar{F}^2_f(M^2)$ in the bifurcation plays the role of
 $F^{\infty2}_f (:= F^2_f(\Lambda^2=\infty)$)
 in the full nonlinear numerical analysis in Section 4.
\par
Let us now consider the  curve (RG flow) on which
both the decay constant $F_\pi$ and
the fermion mass $M :=B(M^2)$ are fixed,
in the three-dimensional parameter space $( h, \lambda ,Z_\phi)$.
There are three types of flows corresponding to the sign of
$F_\pi^2 - \bar{F}_f^2(M^2)$.
For the case (II)
$F_\pi^2 > \bar{F}_f^2(M^2)$ we have $Z_\phi \to  \infty$,
because the condensation function $V(\Lambda^2)$ goes to zero
as $\xi_f := \Lambda /M \to \infty$.
In the case (I)  $F_\pi^2 < \bar{F}_f^2(M^2)$
we can have $Z_\phi = 0$.
This is consistent with the numerical result.
\par
However,
for $F_\pi^2 = \bar{F}_f^2(M^2)$
we obtain a finite value for $Z_\phi$,
since the first term of Eq.\Ref{eq:Zphi} is identically zero
and the second term converges
according to the asymptotic form
 of $B(x)$ and $V(x)$, Eqs.\Ref{eq:asyformZs1}--\Ref{eq:asyformZl2}.
 Thus the  surface $F_\pi^2 = \bar{F}_f^2(M^2)$ yields
renormalized trajectories.
Then we calculate the fixed line, i.e.,
the intersection of the surface of the renormalized trajectories
$F_\pi^2 = \bar{F}_f^2(M^2)$
and the critical surface Eq.\Ref{eq:critical2}.
On the critical surface ($\Lambda/M \to \infty$),
the renormalized trajectories (Eq.\Ref{eq:Zphi} for
$F_\pi^2 = \bar{F}_f^{2}(M^2)$) take the form
\begin{eqnarray}
    Z_\phi &=&
     r_\phi^2 \frac{\pi^2}{2N_c}
       \frac{ \frac{d}{dx} \bar{F}_f^{2}(x) }
            {  h \frac{d}{dx}V(x)^2 }
       \Bigr |_{x = \Lambda^2 \to \infty} \nonumber\\
    &=&
      - {r_\phi^2 \over 16h} \frac{
         B(x)\left[
           B(x)+\frac12 \left( \frac{Z_\phi h/N_c}{(r_\phi+Z_\phi)^2}
               +\lambda \right)V(x)
             \right]
      }
      {V(x)[B(x)-V(x)]} \Bigr |_{x = \Lambda^2 \to \infty},
      \label{eq:FpiCritical}
\end{eqnarray}
where the differentiation of $B(x)$ and $V(x)$ are rewritten  by
using the differential equation Eqs.\Ref{eq:dfeqB},\Ref{eq:SDeqB2}.
\par
Substituting  the UVBC \Ref{eq:UVBC} into Eq.\Ref{eq:FpiCritical}
we obtain a relation among the parameters
$( h, \lambda, Z_\phi )$:
\begin{eqnarray}
  &&
   r_\phi^2
    \left( \frac{4Z_\phi/r_\phi+4+1/N_c}{Z_\phi+r_\phi} h
      +\lambda\right)
    \left( \frac{4Z_\phi/r_\phi+4+1/N_c}{Z_\phi+r_\phi} h
           + \frac12 \frac{Z_\phi/N_c}{(Z_\phi+r_\phi)^2 } h
           +\frac32 \lambda
    \right)
\nonumber\\
 && +16hZ_\phi
   \left( \frac{4Z_\phi/r_\phi+4+1/N_c}{Z_\phi+r_\phi} h
     +\lambda-1 \right)
 =0 \; .
    \label{eq:CriticalFpi}
\end{eqnarray}
Note that Eq.\Ref{eq:CriticalFpi}
is not the renormalized trajectories unless the parameters
$(h,\lambda,Z_\phi)$ lie on the critical surface, since $\Lambda/M \to
\infty$ is already taken in Eq.\Ref{eq:FpiCritical}.
By requiring both
Eq.\Ref{eq:CriticalFpi} and Eq.\Ref{eq:critical2} to be satisfied
simultaneously, we finally obtain the fixed line
which is shown in Fig. \ref{fig:critical}.
Specifically, we find that the fixed line extends
from ($h,\lambda,Z_\phi$) = $(-\frac{1}{16+4/N_c},1/4,0)$
to $(1/4,0,\infty)$
when $r_\phi=1$.
This is one of the main results of  this paper.
\par

Now we state the result on the fixed line.
The fixed line on the critical surface separates
$F_\pi = \bar{F}_f(M^2)$ surface into two regions, one of
which extends to  $Z_\phi \rightarrow \infty$ as
$\Lambda /M \rightarrow \infty$
and another
to $Z_\phi \rightarrow 0$ for $\Lambda/M \gg 1$.
The surface $F_\pi= \bar{F}_f(M^2)$ may be identified
with the surface of renormalized trajectories.
So we can imagine the renormalized trajectories lying on the surface
$F_\pi^2 = \bar{F}_f^2(M^2)$ terminate at the fixed line
on the critical surface.

This picture is consistent with the numerical result in Section 4.
Thus the analytical study confirm the existence of
the renormalized trajectories and the fixed line.
It is again emphasize that $Z_\phi \to \infty$ as $\lambda \to 0$,
i.e.,  no (nontrivial) fixed point in the pure Yukawa model.
 The fixed line is only revealed by presence of the gauge coupling.


\section{Conclusion and discussion}
 In this paper we have investigated phase structure of the
 ($SU(N_c)$-, $U(1)-$) gauged
Yukawa model with a global symmetry  $SU(2)_L\times SU(2)_R$,
using the ladder SD equation with ``standing'' gauge coupling,
Eq.\Ref{eq:SDeq}, and the newly derived
``generalized Pagels--Stokar (PS) formula''
Eqs.\Ref{eq:Fpi}--\Ref{eq:Fpif}.
In  the three-dimensional space of bare parameters
($h$, $\lambda$, $Z_\phi$),  we have obtained the $\xi_f$-constant
surface ($\xi_f := \Lambda/M$, fermion correlation length),
or the scaling relation
(see Fig.\ref{fig:length}, Eq.\Ref{eq:scaling}).
Then, as a limit of $\xi_f \rightarrow \infty$ or the
fermion dynamical mass $M \rightarrow 0$ ($\Lambda$-fixed),
we obtained the critical surface (Fig.\ref{fig:length},
Eq.\Ref{eq:critical2})
 which separates the spontaneous chiral symmetry breaking phase
($M \ne 0$) and  the chiral symmetric phase ($M = 0$).

\par
We have further obtained the ``renormalization-group (RG)
flows'' by requiring
the NG boson decay constant $F_\pi$ as well as $M$
be fixed as we change the cutoff
$\Lambda$ (Figs.\ref{fig:flow},\ref{fig:2dimflow}).
In the three-dimensional space each RG flow is a
cross section of the equi-$F_\pi$ surface sliced
by the $\lambda$-constant plane.
 Among such RG flows we have discovered
 the ``renormalized trajectories'' which terminate on the
critical surface
(Figs.\ref{fig:flow},\ref{fig:2dimflow}).
  The intersection of  a set of renormalized trajectories
  with the critical surface
  may be identified with  a set of UV fixed points,
  the UV fixed line (Fig.\ref{fig:critical}),
 which
exists only for nonzero gauge coupling ($\lambda>0$).
In the $N_c \to \infty$ limit we obtained an explicit analytical
form of the renormalized trajectory as well as the fixed
line, Eqs.\Ref{eq:equxi1/N}--\Ref{eq:rentrj1/N}.
The analytical study has shown that the qualitative feature of
 the phase structure does not depend on $N_c$.
The fixed line extends connecting
$(h,\lambda, Z_\phi)$ = $(-\frac{1}{16+4/N_c},1/4,0)$ and
 $( 1/4, 0, \infty)$.
We thus have established
existence of the fixed line in the gauged Yukawa model ($\lambda>0$)
and non-existence of the (nontrivial)
fixed point in the pure Yukawa model ($\lambda=0$)
through the analytical study and the numerical one within
the framework of the ladder SD equation. This strongly suggests that
although the pure Yukawa model might be ``trivial'',
the gauged Yukawa
model may be a ``nontrivial'' (interacting) continuum field theory
thanks to the presence of the gauge coupling.
\par
The above fixed line should be compared with that
in the gauged NJL model.
 The $Z_\phi \rightarrow 0$ limit of the
gauged Yukawa model is equivalent to the gauged NJL model with
 appropriate rescaling of the Yukawa coupling (see Eq.\Ref{eq:gNJL}).
Now in the gauged NJL model with the standing gauge coupling,
the fixed line is identified with the critical line
itself\cite{kn:KMY89,kn:NSY89,kn:KTY93a}, while
the fixed line in the gauged Yukawa
model deviates from that of the gauged
NJL model ($Z_\phi=0$) and drifts into the direction of $Z_\phi>0$
on the critical
surface in the three-dimensional bare parameter
space (Fig.\ref{fig:critical}).
Actually, the projection
of the fixed line onto the  $Z_\phi=0$ plane exactly coincides
with the
critical line of the gauged NJL model.
 This result is due to the
propagating (``hopping'') degree of freedom of the scalar field,
since the
corresponding scalar field in the gauged NJL model is merely
an auxiliary one
and has no kinetic term.
\par
It was pointed out\cite{kn:KSY91} that, within the gauged NJL model,
the line of $M$-constant does not exactly coincide with that of
$F_\pi$-constant particularly in the weak gauge coupling region.
 Actually, $M$ is a slowly decreasing function of the cutoff
$\Lambda$ along the $F_\pi$-constant line as was demonstrated
in the top
quark condensate model\cite{kn:MTY89,kn:Yama92}.\footnote{
This is not a problem of the top quark condensate model
as a phenomenological
model with an explicit cutoff. This in fact was the
very predictability of the top quark mass in that model.
Actually, we can even take the formal limit $\Lambda \to \infty$
of the model and obtain a finite (``nontrivial'') continuum
theory\cite{kn:KSY91,kn:Yama92},
although such a limit is peculiar in the sense we just mentioned.
}
This suggests that we need (at least one) other relevant operators
 to obtain a  sensible continuum field theory. Having introduced
 a $Z_\phi$ degree of freedom, we can explicitly see how this problem
 is resolved in this enlarged coupling space.
Let us discuss the small $Z_\phi(>0)$ in the region (I):
$F_\pi<F_f^\infty$, which is bounded between the surface
of renormalized trajectories ending
at the fixed line ($F_\pi=F_f^\infty$
surface) and the $Z_\phi=0$ plane (see Fig.\ref{fig:2dimflow}).
 If we specify a point on the
RG flow
and perform one step of RG (by decreasing the cutoff
to the smaller $\xi_f$),
 the point moves to
another point on the same flow with a larger $Z_\phi$.
 Thus the requirement of both $F_\pi$ and $M$ to be
constant can only be met by the change in $Z_\phi$ direction
and the continuum theory
is obtained at the fixed line with $Z_\phi>0$.

\par
Here we should remark that our results (the existence of the
critical surface, the fixed line and the renormalized trajectory) are
based on the ``standing'' ansatz of the gauge coupling.
The RG evolution of the scalar quartic
coupling $\lambda_\phi$ is also disregarded in the present paper.
In principle, it is possible to take these effects into account in our
analysis by fixing yet another ``physical quantities''.
The mass of the physical scalar boson $m_\phi^{\rm phys}$ and the
(infrared) scale of $SU(N_c)$ gauge interaction $\Lambda_{\rm gauge}$
would suit such a purpose.
We then need to estimate the possible feedback from  the running
of those parameters to the restricted analysis here
within the essentially
two parameter space $(h,Z_\phi)$.

As for the running of $\lambda_\phi$, we need to
modify two aspects  so as to take account of the feedback in the SD
equation utilized in this paper.
One is the inter-relation among $\langle \bar\psi\psi\rangle$ and
$\langle\sigma\rangle$
which should be changed with respect to the quartic
coupling $\lambda_\phi$ (see Eq.\Ref{eq:oderpara}).
This effect is, however, suppressed by
$\langle\bar\psi\psi\rangle/\Lambda^3$ and
thus we hope can be
disregarded near the critical point as far as the phase
transition is the
second order.
   Another is the use of
the $\lambda_\phi$ dependent dressed propagators of the
scalar/pseudoscalar field
in the  rainbow graph of the scalar/pseudoscalar.
This is also suppressed by $1/N_c$ and is expected to give only
a tiny effect for large $N_c$.

It is a rather delicate problem, however, to estimate the feedback
from
the (asymptotically free) running of the gauge coupling $\lambda$.
It is straightforward to include such
a running effect (at one-loop) into
our analysis as we did
in the gauged NJL model ($Z_\phi=0$)\cite{kn:KSY91}
where almost all the RG flows tend to the pure NJL point
$(h,\lambda)=(h_c,0)$ as $\Lambda \to \infty$; namely, the fixed line
shrinks into a single point $(h,\lambda)=(h_c,0)$, the UV fixed point,
due to the one-loop running effect of
the gauge coupling\cite{kn:KSY91}.\footnote{
In Ref.\cite{kn:KSY91} we found another UV fixed point
$(h,\lambda)=(0,0)$, the
``pure QCD point'', which has no correspondence
to the fixed point of the
model with the standing gauge coupling discussed in this paper.
}
Thus the fixed line of the gauged Yukawa model
in $(h,\lambda,Z_\phi)$ space may also shrink into $\lambda \to 0$
for the running
gauge coupling, since its projection onto the $(h,\lambda)$ plane
is again expected to be that of the gauged NJL model as in the
case of the standing gauge coupling discussed above.
    Now, this limit is rather subtle, since
the fixed line extends
into the region $Z_\phi\rightarrow\infty$ as $\lambda\rightarrow0$,
see Fig.\ref{fig:critical}.
Thus even the qualitative structure of the RG seems
to depend critically
on the speed of running of the gauge coupling,
or on the number of quark
flavors for the QCD coupling.\cite{kn:KSY91}
Similarly, in the perturbative RG analysis of the standard
model\cite{kn:PL81}, the qualitative structure
of the RG flow of the Yukawa coupling in fact depends
on the speed of the running of the QCD coupling.
This problem will be dealt with in the forthcoming paper.
\par
The phase structure obtained in this paper may be compared
with results of
the lattice theory.  If we restrict the model to the pure Yukawa case
($\lambda=0$),  our results on the phase structure
are consistent with those
of the lattice Yukawa model \cite{kn:BJJN89}.
We here predict the phase structure of the gauged Yukawa model on the
lattice using the parameterization which is similar
to the lattice theory.
After the transformation of the parameters ($d$: dimensionality of
the space-time, $a$: lattice spacing):
\begin{equation}
 \kappa_H = \frac{4 Z_\phi}{2d Z_\phi + a^2 M_\phi^2},
 \qquad
  g_Y^2 = \frac{1}{N_c} \frac{g_y^2}{2d Z_\phi + a^2 M_\phi^2},
\label{eq:laticeparm}
\end{equation}
our Yukawa model (excluding the gauge part)
can be cast into the lattice
version with the lattice action:
\begin{eqnarray}
  S &=&  -\kappa_H \sum_{x,\mu} a^{d-2}
     \varphi_\alpha(x) [\varphi_\alpha (x+\mu)+\varphi_\alpha(x-\mu)]
         +4\sum_{x} a^{d-2} \varphi_\alpha^2(x)
    \nonumber\\
    & & +\frac{1}{2} \sum_{x,z,\mu} a^{d-1}
         \bar \psi_i(x) \gamma_\mu[\delta_{x+\mu,z}-\delta_{x-\mu,z}]
         \psi_i(z)
    \nonumber\\
    & & +\sqrt{8} g_Y \sum_{x} a^d
         \bar \psi_i(x) [
          \varphi_0(x) +i \gamma_5 \vec{\tau} \cdot \vec{\varphi}(x)]
         \psi_i(x),
\end{eqnarray}
with $i=1,..,N_c$ and
$\alpha=0,..,3$ being indices of the $SU(N_c)$ fundamental
representation and the vector representation of $O(4)=SU(2)_L\times
SU(2)_R$, respectively.
 The phase structure of the gauged Yukawa model as well as the pure
 Yukawa model in this parameterization is shown in Fig.\ref{fig:latt}.
Our claim on the existence of the
fixed line in the presence of the gauge coupling is based on the
nonperturbative analysis through the SD equation.
Actually, such a remarkable
feature does not appear by perturbatively including the effect of
the gauge
coupling \cite{kn:GPS92}.
  In view of this  there exists, to our knowledge,
  no available data from the lattice
theory which are  comparable with our result.
Our finding in this paper will be sufficient
 to urge the lattice people to
try to perform a full study of the gauged Yukawa model on the lattice,
which will really confirm whether our claim is right or not.

\par
In this paper we have assumed $Z_\phi >0$, since otherwise the vacuum
would be unstable already in the scalar/pseudoscalar sector
(there appear tachyon ghosts).
In fact the ladder SD equation
has a tachyon pole and cannot be Wick-rotated for
$Z_\phi < - r_\phi$. Furthermore the RG flow becomes singular at
$
Z_\phi= -r_\phi[1 - (\sqrt{4N_c +1} - 1)/(4N_c)]
$
$(>-r_\phi) $, which can be seen from the
 UVBC Eq.\Ref{eq:UVBC} and the generalized Pagels-Stokar
  formula Eqs.\Ref{eq:Fpi}--\Ref{eq:Fpif}.
However, we can formally extend our RG flow to the
negative value
$
 0 > Z_\phi >
  -r_\phi[1 - (\sqrt{4N_c +1} - 1)/(4N_c)]
$
in our analysis.
Although this $Z_\phi$ region might also be pathological,
it is amusing to compare our RG flow with the lattice Yukawa model.
 By the above correspondence we can extend the RG flow of Fig.10a
 (no gauge
 coupling) into the negative $\kappa_H$ region, which indicates no sign
 of fixed point. This is consistent with the recent lattice
  analysis of the pure Yukawa model.

\par
Finally, besides the RG analysis,
we consider a physical application of the
above information on the phase structure.
Here we keep the UV cutoff $\Lambda$ fixed. This is actually the
case for the Yukawa-driven top mode
model\cite{kn:KTY90,kn:CR91} and the heavy scalar technicolor
model\cite{kn:Simm89}.
First we discuss the criterion
whether the symmetry breaking is ``dynamical'' or not.
This is not so clear
in the system having the elementary scalar from the onset,
since it always
mixes with the composite one. However, as we can see from our
phase diagram, we can discriminate two distinct
region separated by the renormalized trajectories:
region (I) ($F_\pi<F_f^\infty$) and region (II) ($F_\pi>F_f^\infty$).
In region (I) we can always arrange the fermionic contribution
$F_f^2$ to
dominate the bosonic one $F_b^2$ in the the NG boson decay constant
$F_\pi^2$, particularly for a small $Z_\phi$ or a large $\Lambda$
(see Fig.~\ref{fig:ratio}).
This is contrasted with the region (II) where $Z_\phi$ is relatively
large and the bosonic contribution persists even for a large $\Lambda$.
We may call such a fermionic dominance of the $F_\phi$ in region (I)
the ``dynamical'', in contrast with the fermionic non-dominance
in region (II)
which may be called the ``non-dynamical''. Although the nomenclature
is somewhat obscure, our definition is based on the phase diagram and
hence is without ambiguity. This concept may be useful for analyzing
the unified models based on the strong fermion dynamics involving
elementary scalar field.

\par
Next we discuss the dependence of the
fermion dynamical mass $M$ on $Z_\phi$.
 For a given value of $F_\pi$, we have shown that $M$ decreases as
$Z_\phi$ increases along the $\Lambda$-fixed line,
see Fig. \ref{fig:ratio}.
This implies that $M$ in the gauged Yukawa model ($Z_\phi >0$)
is always
smaller than that in the gauged NJL model ($Z_\phi =0$) for the same
$F_\pi$ value, which is
actually derived more generally as a direct consequence of our
generalized PS formula Eqs.\Ref{eq:Fpi}--\Ref{eq:Fpif}
as was explained in Section 3.
This might appear rather peculiar from the general view point of a
 ``generalized NJL model'' (without gauge coupling)
 of Hasenfratz et al.\cite{kn:HHJKS91} presuming
  arbitrary higher dimensional operators.
  Such operators  yield a priori an equal potentiality
 either to raise or lower the fermion dynamical mass compared with the
 NJL model. On the contrary, we have specified a
 model, the gauged Yukawa model, for the higher dimensional operators
  instead of introducing arbitrary such operators.
Therefore we have obtained a definite answer that the fermion mass in
this model must be lower than in the gauged NJL model.
If this conclusion in the $SU(2)_L \times SU(2)_R$-invariant gauged
Yukawa model persists in the $SU(2)_L \times U(1)_Y$-invariant
gauged Yukawa model, then the Yukawa-driven top quark condensate
based on the latter
model will give a smaller top quark mass prediction than the original
one\cite{kn:MTY89,kn:BHL90} based
on the $SU(2)_L \times U(1)_Y$-invariant
gauged NJL model.  The analysis along this direction including the
running effects of QCD coupling is under way.

{\bf Acknowledgements}
\par
  We would like to thank I. Tkachev
for stimulating and valuable discussions.
Thanks are also due to W.A. Bardeen, B. Holdom and V.A. Miransky
for discussions.
K.Y. thanks the Aspen Center for Physics for hospitality while
some of this work was done.

\newpage

{\large \bf Figure Captions}
\begin{figure}[h]
  \caption{\hspace{\textwidth}}
  \label{fig:Fpi}
  Graphical representation for the NG boson decay constant $F_\pi$
  \protect Eq.\Ref{eq:fpiform}.
\par

\caption{\hspace{\textwidth}}
\label{fig:yukawa}
Critical line of the pure Yukawa model ($\lambda =0$).
The critical line is drawn by the bold line and
the solid lines denote the renormalization-group flows.
Each (upper) flow corresponds to
a different (larger) value of $F_\pi$.
The dotted lines are the equi-$\xi_f$ (correlation length) lines.

\caption{\hspace{\textwidth}}
\label{fig:orderparameter}
  Yukawa coupling dependence of the chiral condensate.
The solid line, broken line, dashed line and dotted
line correspond to $Z_\phi=0.01,0.1,1,10$, respectively.

\caption{\hspace{\textwidth}}
\label{fig:gNJL}
Critical line (bold line) and equi-$\xi_f$ lines
for $Z_\phi=0$ (gauged NJL model).

\caption{\hspace{\textwidth}}
     Equi-$\xi_f$ (correlation length) surface in the space
  ($h$, $\lambda$, $Z_\phi$). The framed surfaces from top
  to bottom corresponds to $\xi_f \equiv \Lambda/M=\exp(t/2)$
  for $t=$ 1,3,5,7,15. The final surface is very close
  to the critical surface ($\xi_f=\infty$).
\label{fig:length}

\caption{\hspace{\textwidth}}
   Renormalization-group flows for fixed $F_\pi$,
  which are  shown as
  the intersection of the equi-$F_\pi$ surface
  with the fixed-$\lambda$ (gauge coupling ) plane.
(a) $F_\pi^2/M^2$ = 0.142,
(b) $F_\pi^2/M^2$ = 0.086.
\label{fig:flow}
\end{figure}
\vfill \eject

\begin{figure}[t]
\caption{\hspace{\textwidth}}
Critical line (bold line) and renormalization-group
 flows (solid line) for the gauged Yukawa model ($\lambda >0$).
They are depicted in the fixed-$\lambda$ plane.
Each (upper) flow corresponds to a different (larger)
value of $F_\pi$.
The dotted lines denote the equi-$\xi_f$ lines.
(a) $\lambda=0.05$,
(b) $\lambda=0.15$.
\label{fig:2dimflow}

\caption{\hspace{\textwidth}}
      The ratio $F_b^2/F_f^2$
      along the renormalization-group flow
      in the fixed-$\lambda$ plane.
      The line passing through the origin denotes
      the fixed-cutoff $\Lambda$ line.
(a) $\lambda=0.02$, (b) $\lambda=0.08$.
\label{fig:ratio}

\caption{\hspace{\textwidth}}
   The mesh denoted by sold line shows
    the critical surface obtained from the analytical
   solution of the SD equation Eq.\protect\Ref{eq:SDeqB1}
   and Eq.\protect\Ref{eq:dfeqB}
   in the limit of $\Lambda/M \to \infty$.
   The fixed line on the critical surface is plotted by bold line.
  The projection of the fixed line to
  the $Z_\phi$-$h$ plane (a)
  and the $\lambda$-$Z_\phi$ plane (b)
  are plotted by solid line.
  The equi-$\lambda$ lines (a)
  and equi-$h$ lines (b) are shown
  by the lines corresponding to different values of
  $\lambda$ and $h$ (see the upper right corner), respectively.
\label{fig:critical}

\caption{\hspace{\textwidth}}
        Critical line (bold line) and
        renormalization-group flows (solid line)
        in the fixed-$\lambda$ plane.
        The parameters are translated by using
        Eq.\protect\Ref{eq:laticeparm},
        and we choose the lattice spacing as
        $a^{2} M_\phi^2 = 50$ and $N_c=1$.
        The dotted line denotes the equi-$\xi_f$
        line.
        (a) $\lambda=0$ (pure Yukawa model),
        (b) $\lambda=0.1$.
\label{fig:latt}
\end{figure}
\hspace{10em}

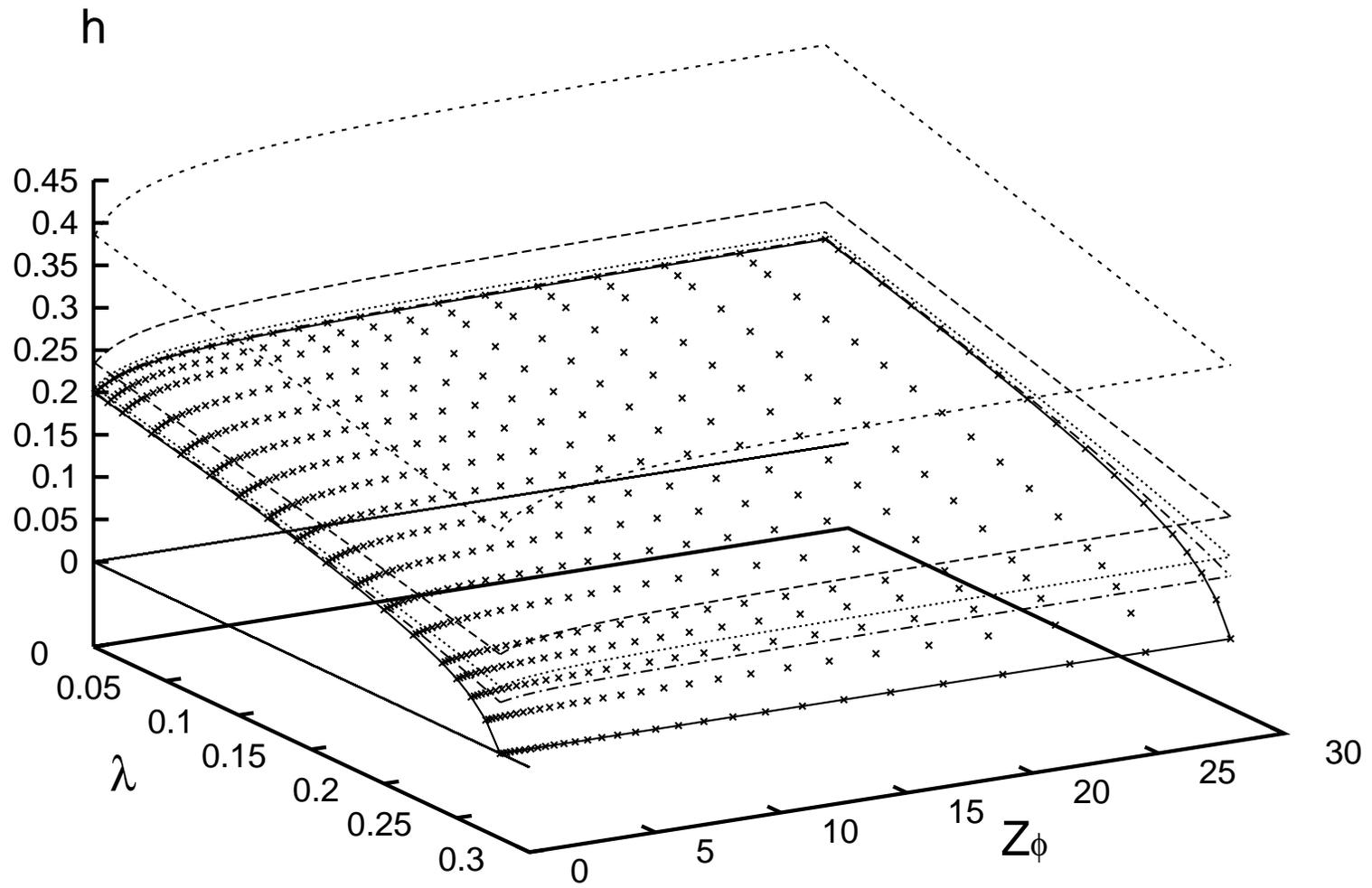
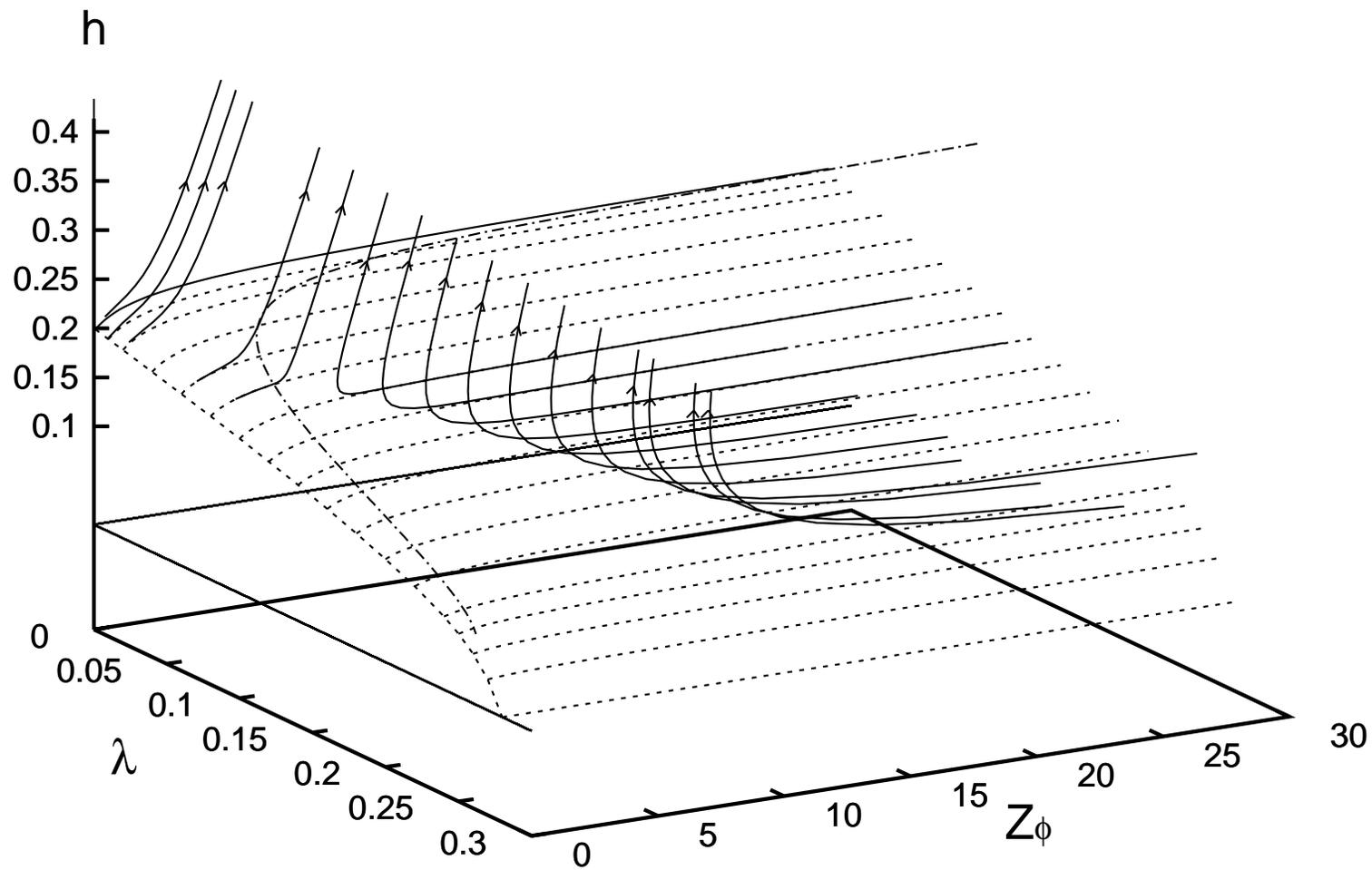
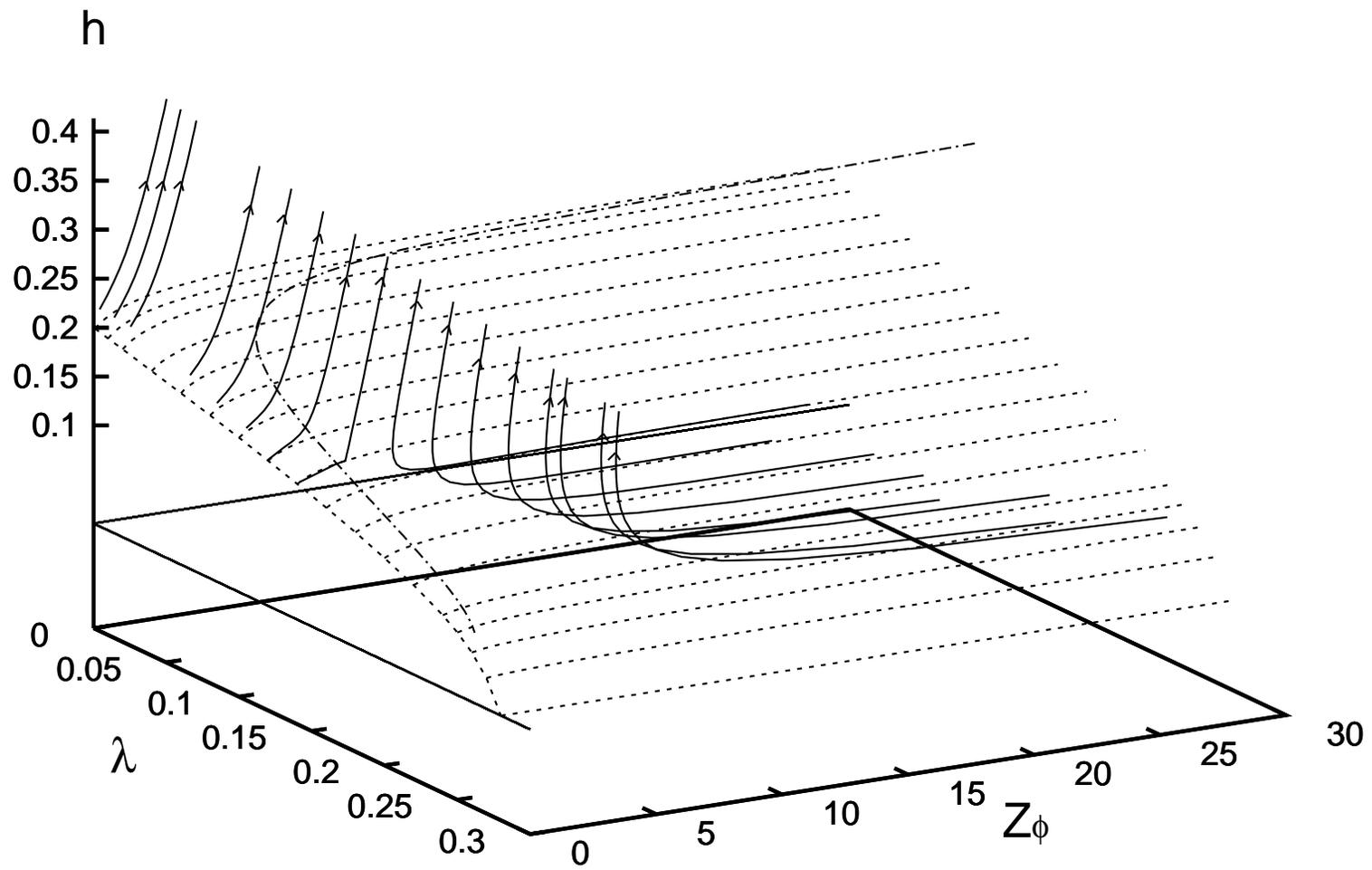
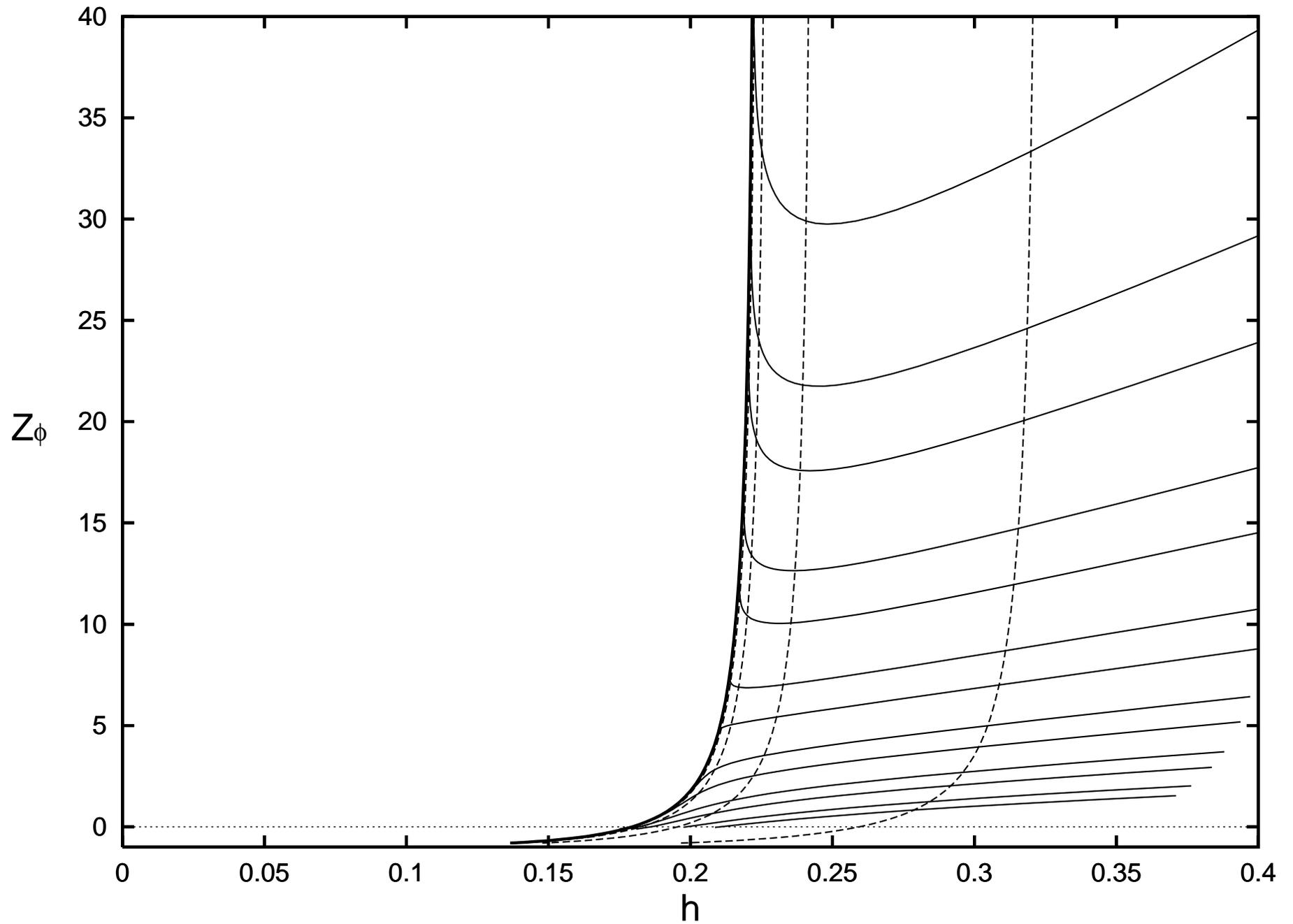
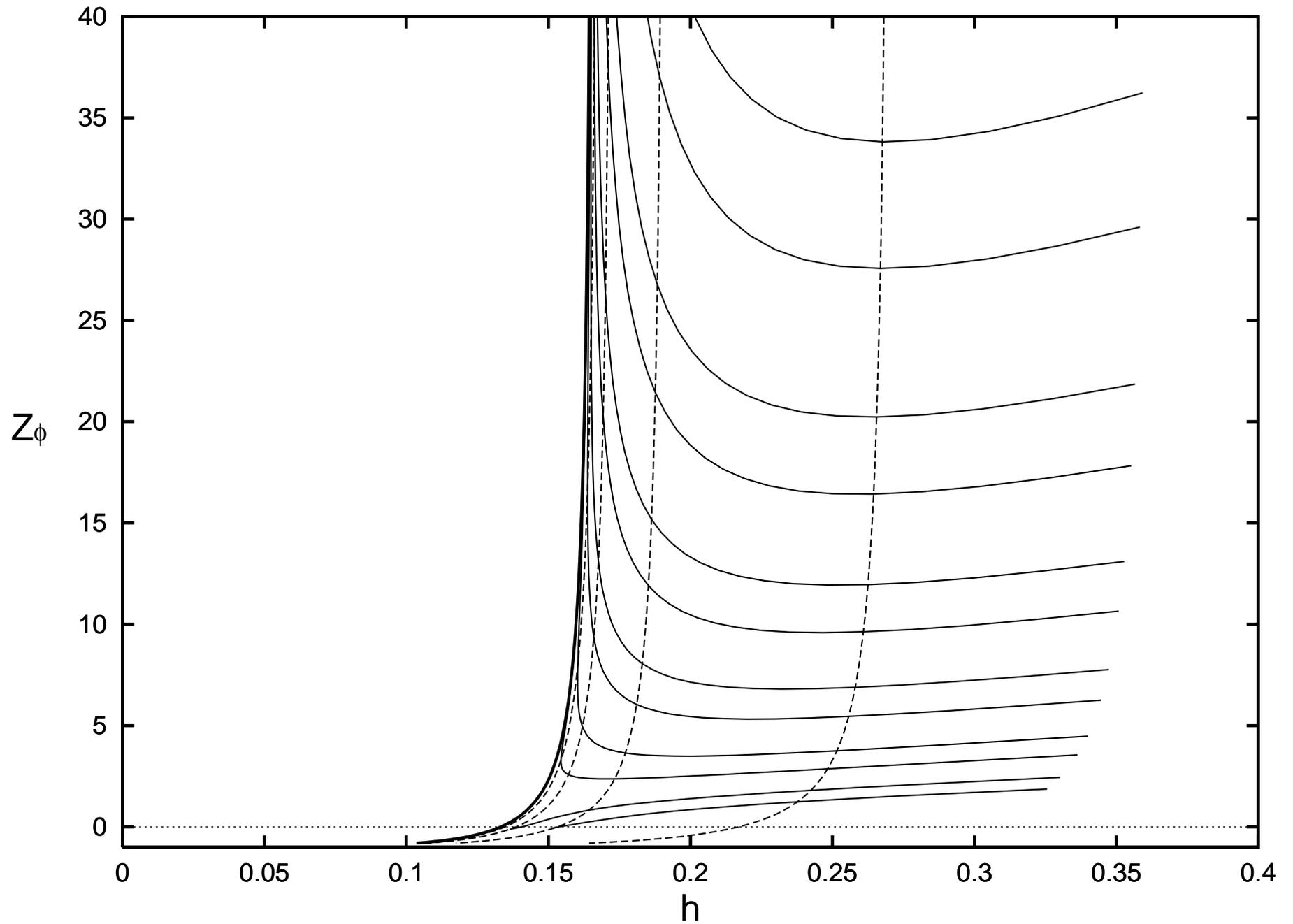
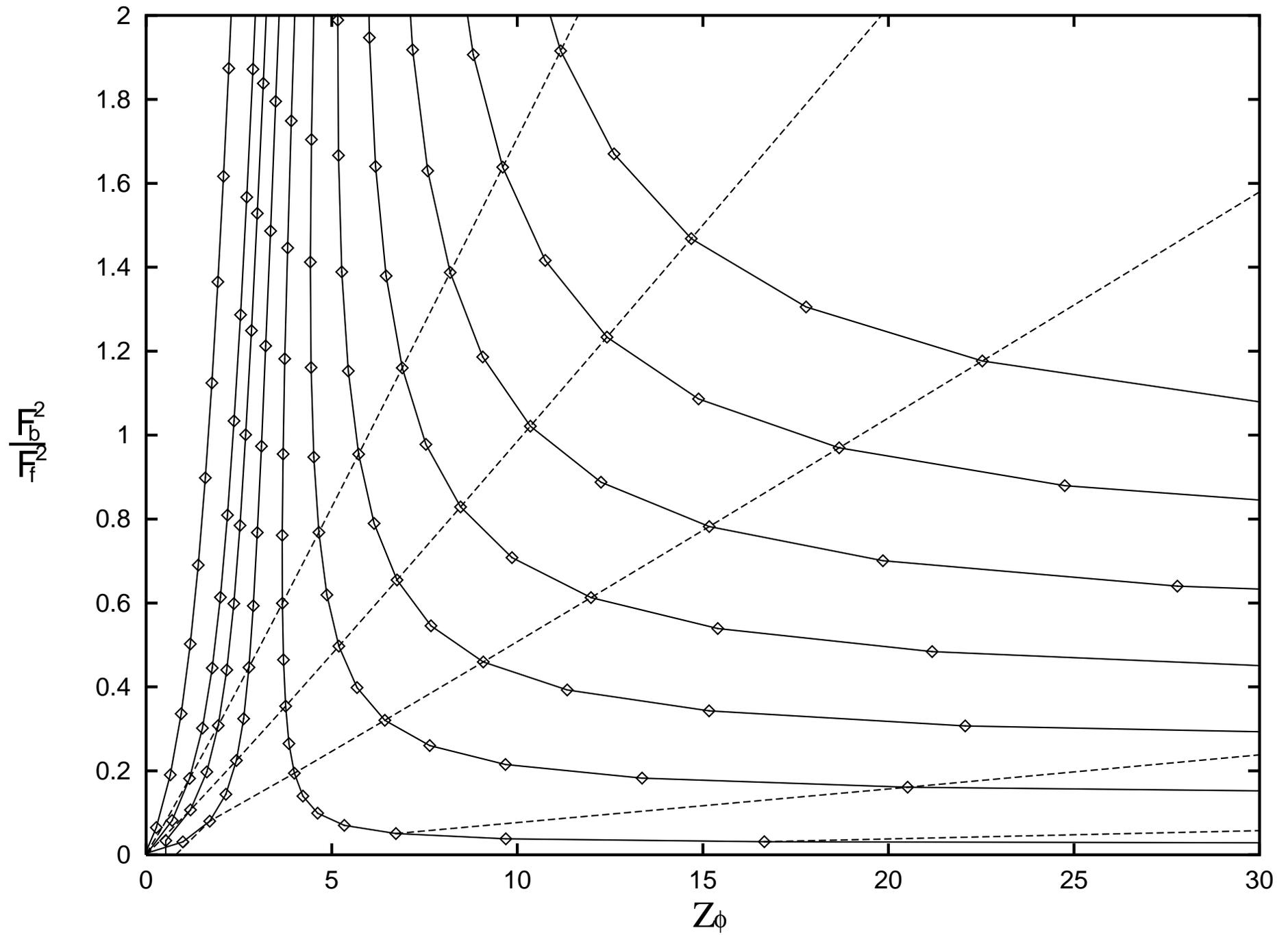
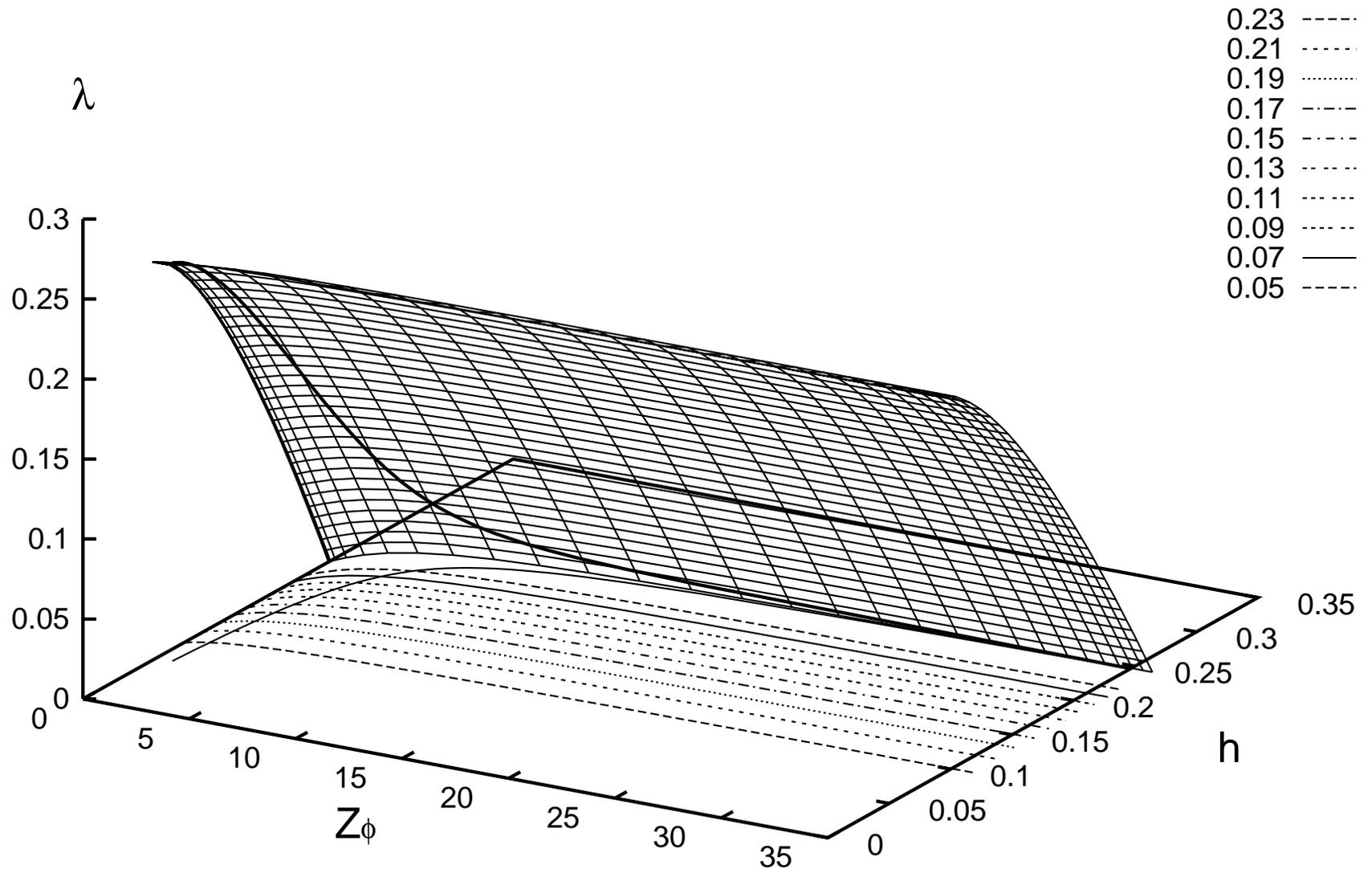
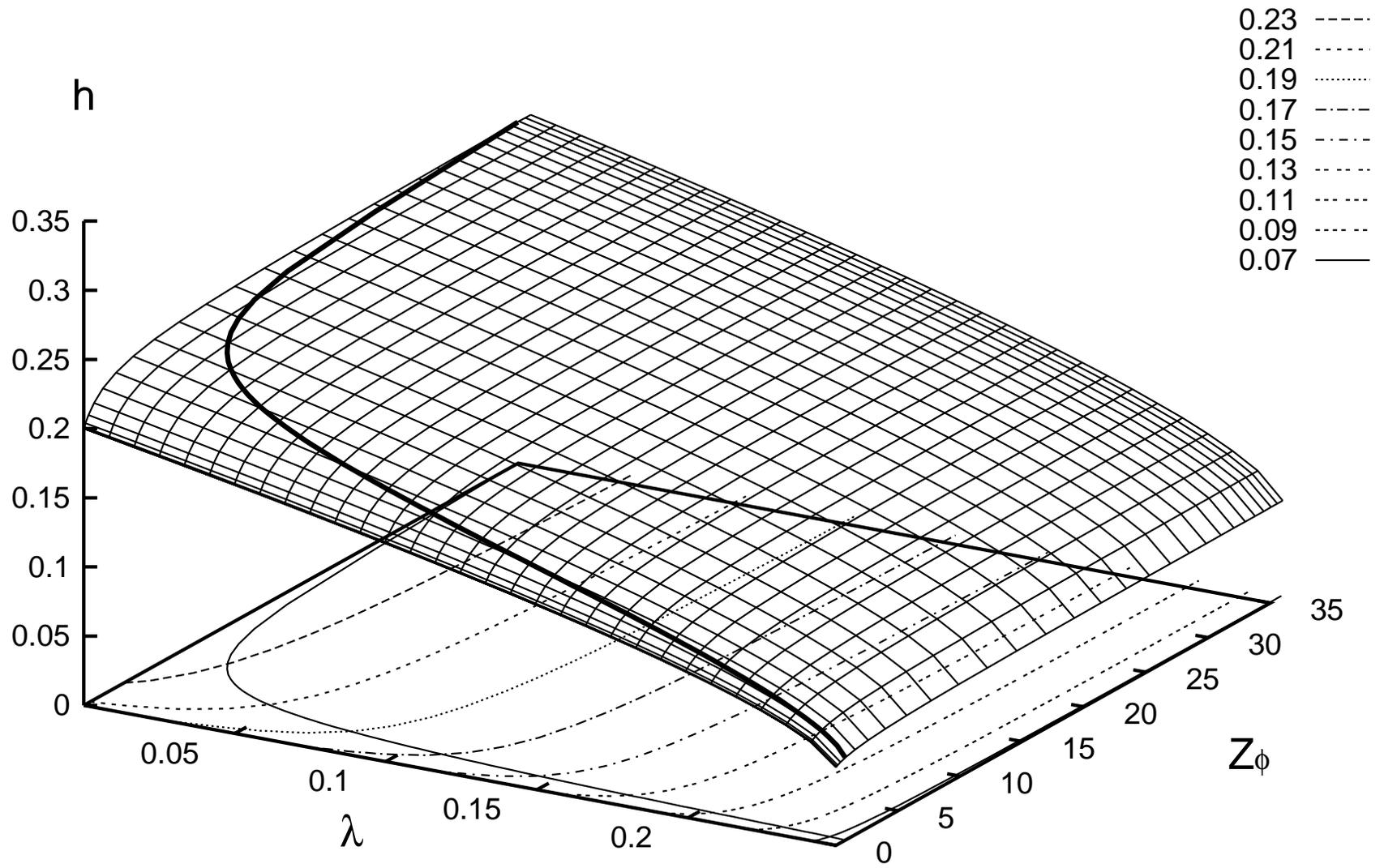
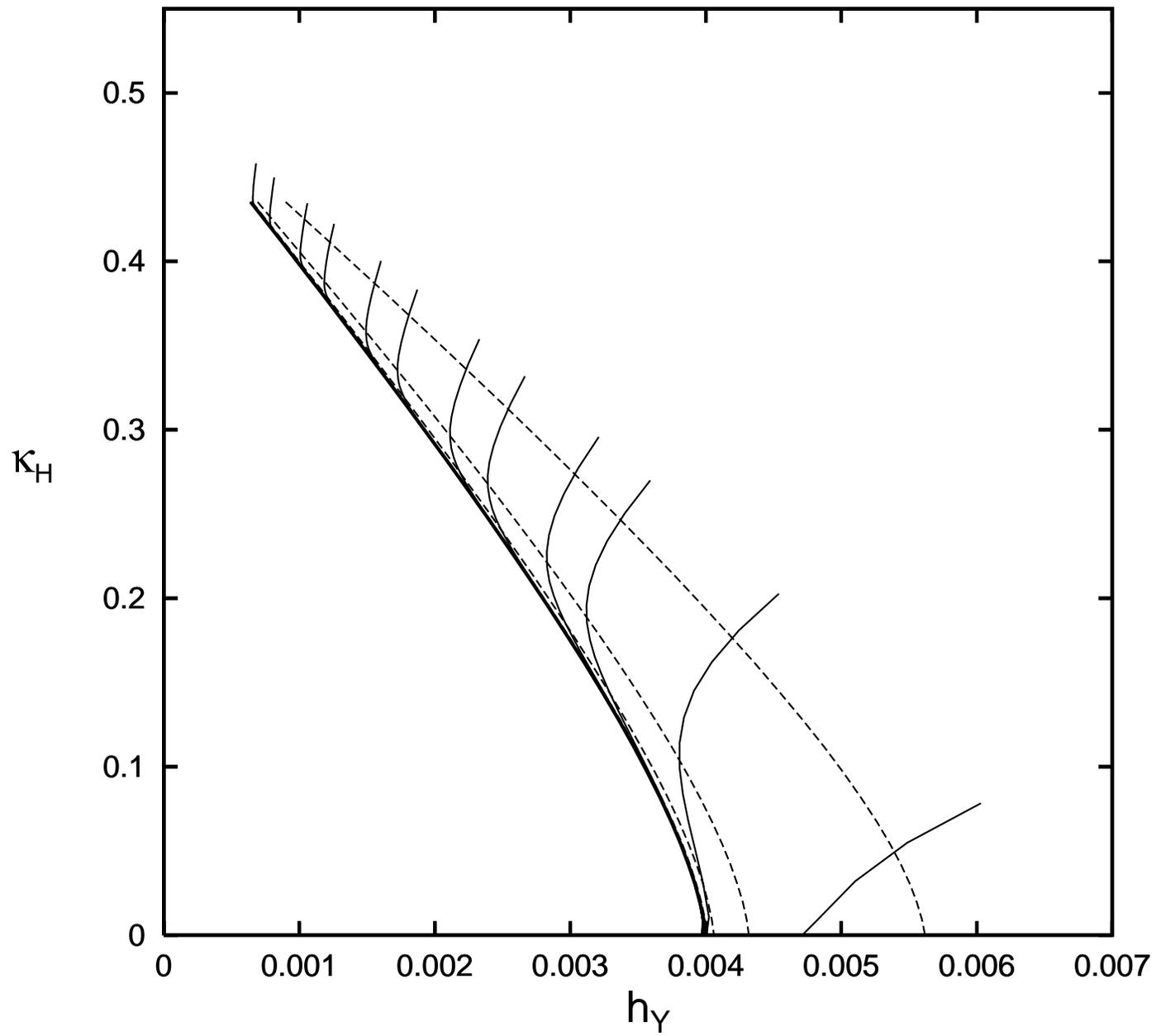
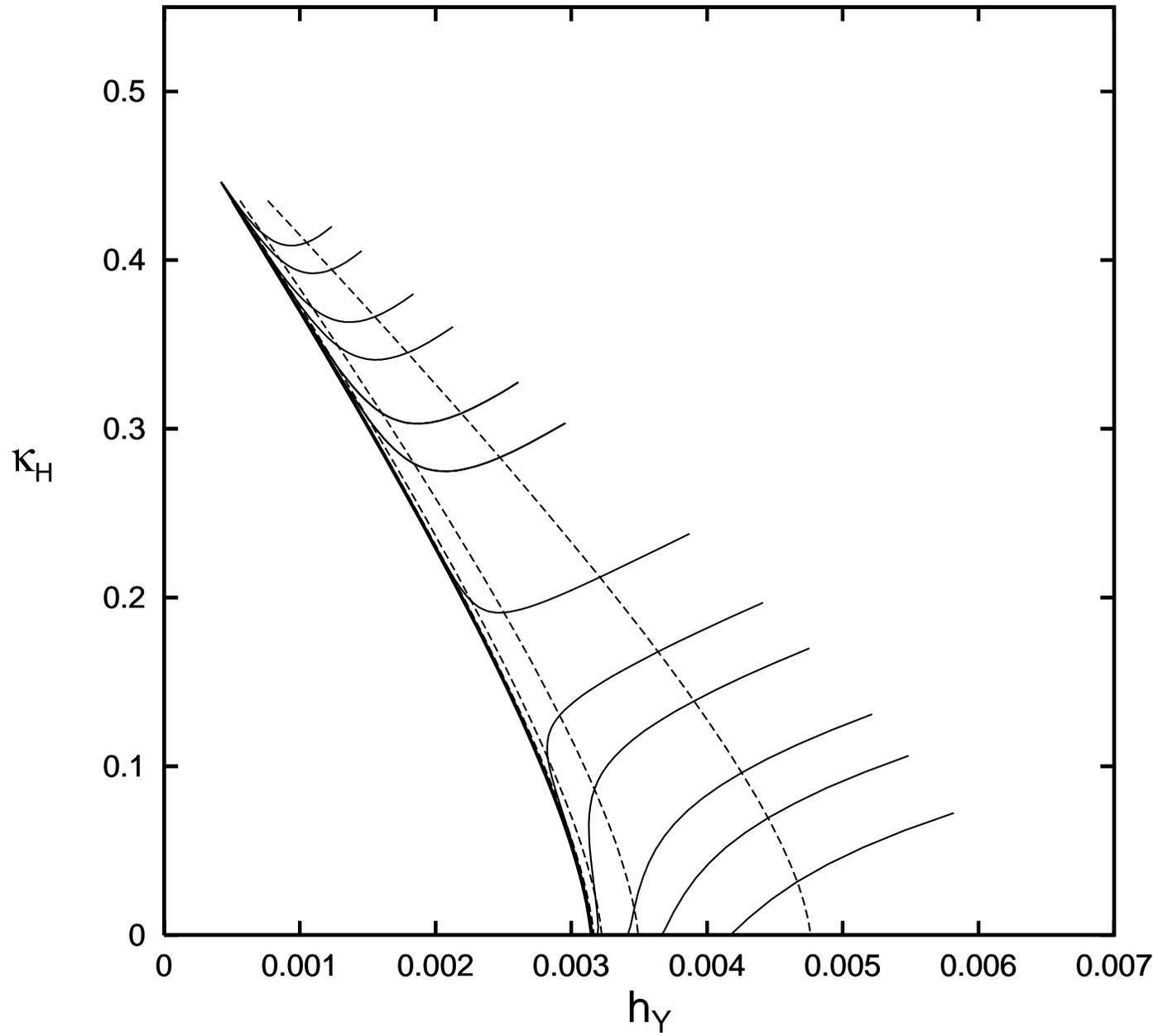
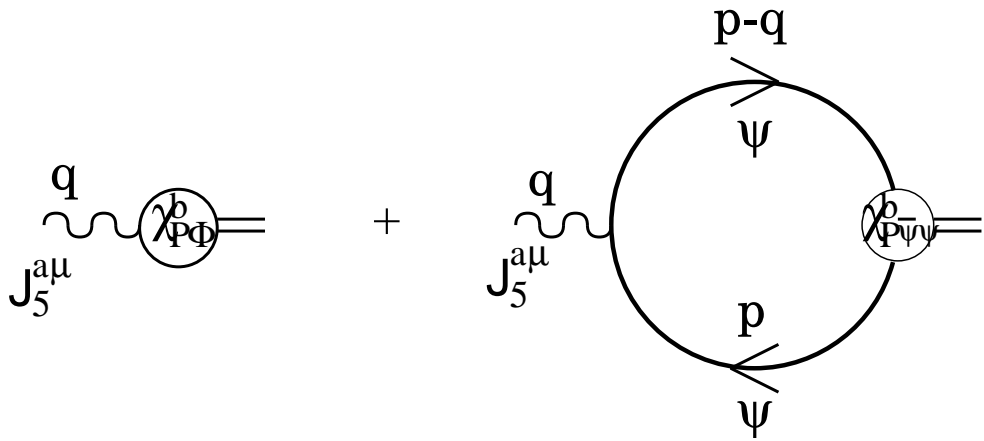
\end{document}
